\documentstyle[12pt]{article}
\def\gev{{\rm \, Ge\kern-0.125em V}}
{}
{}
\def\beq{\begin{equation}}
\def\eeq{\end{equation}}
\newcommand{\beqa}{\begin{eqnarray}}
\newcommand{\eeqa}{\end{eqnarray}}
\newcommand{\beqar}{\begin{eqnarray*}}
\newcommand{\eeqar}{\end{eqnarray*}}

\newcommand{\al}{\alpha}

\renewcommand{\l}{\lambda}

\newcommand{\AH}{S}
\newcommand{\thetn}[1]{\theta^{(#1)}}
\newcommand{\bthetn}[1]{\bar{\theta}^{(#1)}}

\newcommand{\eg}{{\it e.g.,}\ }
\newcommand{\ie}{{\it i.e.,}\ }
\newcommand{\pol}{\varepsilon}
\newcommand{\F}[1]{{[F_{(#1)}]}} 
\newcommand{\T}[1]{{[\theta_{(#1)}]}} 
\newcommand{\Fp}[1]{{[F'_{(#1)}]}} 
\newcommand{\cL}{{\cal L}}
\newcommand{\fun}{{F(\tilde{z},r)}}

\newcommand{\B}{B_0}
\newcommand\e{{\rm e}}

\newcommand\prt{\partial}

\begin{document}
\begin{titlepage}
\pagestyle{empty}
\baselineskip=21pt
\rightline{\small hep-th/yymmddd \hfill McGill 96--44}
\vskip -1em
\rightline{\small WATPHYS-TH-96/17}
\vskip .5in
\begin{center}
{\huge{\bf Wavy Strings: Black or Bright?}}
\vskip 3em

{\large
Nemanja Kaloper,$^{a,}$\footnote{kaloper@avatar.uwaterloo.ca}
Robert C. Myers$^{b,}$\footnote{rcm@hep.physics.mcgill.ca}
and
Harold Roussel$^{b,}$\footnote{roussel@hep.physics.mcgill.ca}}
\vskip 1.5em

{$^a$ {\it Department of Physics,
University of Waterloo, Waterloo, ONT, N3L 3G1 Canada}}
\vskip .5em

{$^{b}$ {\it Department of Physics, McGill University,
Montr\'eal, PQ, H3A 2T8 Canada}}
\vskip 4em

\begin{abstract}
Recent developments in string theory have brought forth
a considerable interest in time-dependent hair on
extended objects. This novel new hair is typically characterized
by a wave profile along the horizon and angular momentum
quantum numbers $l,m$ in the transverse space.
In this work, we present an extensive treatment
of such oscillating black objects, focusing on 
their geometric properties. We first give
a theorem of purely geometric nature, stating that such
wavy hair cannot be detected by any scalar
invariant built out of the curvature and/or matter fields.
However, we show that
the tidal forces detected by an infalling observer diverge at
the `horizon' of a black string superposed with a vibration
in any mode with $l \ge 1$.
The same argument applied to longitudinal ($l=0$) waves detects only
finite tidal forces. We also provide an example
with a manifestly smooth metric, 
proving that at least a certain class of these longitudinal waves
have regular horizons.
\end{abstract}
\end{center}

\end{titlepage}
\baselineskip=18pt

\setcounter{footnote}{0}
\section{Introduction}

One of the most confounding puzzles about black holes in
General Relativity is the apparent incompatibility between
their extremely simple structure, governed by the famous
no-hair uniqueness theorems \cite{Hair},
and their generically large thermodynamic entropy \cite{Entropy}.
The former is indicating that the complexity associated with the latter
is not encoded in the classical solutions.
The crux of the no-hair theorems in General Relativity is
that regardless of how a black hole was formed it
is completely described once its mass, spin and Abelian
electric/magnetic charges are determined. All other types
of long range interactions, such as scalars with or without
self-interactions, massive or non-Abelian gauge fields, etcetera 
are excluded because they would either lead to naked singularities \cite{Hair}
or would be unstable and thus would disappear shortly after
the formation of the black hole \cite{NAb}. More specifically,
the machinery of the no-hair theorems rests on the assumption
of regularity of a stationary null surface, \ie the event horizon,
and the demonstration that it leads to the vanishing of all
matter charges, other than Abelian, which could carry long range
interactions. Because the Equivalence Principle demands that
the fields be coupled minimally to gravity, the vanishing of
charges in stationary spherically symmetric space-times
implies the vanishing of the external fields or hair.
A small deviation from these results are black holes with
non-Abelian hair, where the hair arises because of the
nonlinearities in the gauge field equations of motion; however,
this hair is unstable to perturbations, with the hairy black hole
rapidly decaying into a bald one \cite{NAb}. The final result is that
all of the structure of the configuration which collapsed
into a black hole not encompassed by mass, spin and Abelian
charges must have been radiated away in the process of
formation of the hole.

Recent developments in non-minimal models of gravity, and
especially in string theory, where the modifications of
General Relativity have a firmer foundation, have brought about
some explicit counterexamples to the no-hair theorems \cite{CMP}.
Namely, the models considered involve new non-minimal
couplings which provide extra sources in the equations
of motion. However, whereas the particular solutions to these
modified equations carry
non-trivial long-range hair exterior to black
holes, there are no new constants of integrations and
so these aberrations of the no-hair theorems
are deviations in letter only and not in spirit. 
The new charges are completely
determined by the charges already present in Einstein's theory,
and the non-minimal hair is secondary (as opposed to the primary
hair carried directly by the black hole charges).\footnote{The
secondary hair will be stable because the monopole
charges do not vanish, in contrast to the non-Abelian hair
mentioned above.}
Nevertheless,
the secondary hair does affect thermodynamic properties of
black holes, changing the expressions for both the temperature
and the entropy \cite{CMP}.

Another kind of black hole hair may appear if we abandon the
constraint of stationarity \cite{CMP2}--\cite{Tsey}. 
Recent developments in string theory
have brought forth a considerable interest in time-dependent hair on
black holes and extended objects. In particular it has been suggested
that such hair might give a classical accounting of the black hole
entropy \cite{LW,Tsey,CT}. In any event,
one might expect that this situation would
be a more realistic description of a black hole,
because stationary black holes could exist only as hermits in
complete solitude; yet their very nature precludes this, as their
own gravitational field permeates the whole Universe and
communicates with all of its inhabitants. The principal
obstacle to studying non-stationary problems is an enormous
complexity of gravitational equations of motion, often
resulting in their intractability. Indeed, there are
comparatively few exact non-stationary localized solutions
known to date, and none of them corresponding in all important
aspects to the physical picture of black holes.
Still, some of these solutions may provide useful models to study
realistic hairy black holes.  Most of these solutions
arise because we can embed four-dimensional black holes in higher dimensions,
via a procedure inverse to dimensional reduction (thus, the term oxidation
--- see \eg \cite{rusty}).
In certain special cases, the resulting oxidized solution has a null,
hypersurface-orthogonal, Killing vector \cite{Kall}--\cite{Behr}. 
They can be used as a natural starting point
for the construction of a more general family of non-stationary solutions
characterized by a set of arbitrary functions. The presence
of this isometry indicates that there exists a much larger
family of wave-like solutions which reduces to the black hole-like
solution for the special choice of its degrees of freedom.
A similar correspondence exists, for example, between the
Brinkmann wave solutions in General Relativity and the flat space.
We can get a glimpse at this larger family by using a
solution-generating technique
defined by Garfinkle and Vachaspati \cite{GV} as a method to restore
some of the original wave degrees of freedom.
The resulting solutions still possess a null hypersurface
orthogonal Killing vector, and thus are
describing a gravitational disturbance propagating through the
original environment at a speed of light \ie
a gravitational wave. 

A fundamental question of interest  is to determine if
this larger family still has a regular null surface, which can be reached by
causal or null geodesics, \ie an event horizon.
In this paper, we will examine this question for
a certain family of five-dimensional black strings with 
a single wave profile function in an $(l,m)$ multipole mode.
First, however, we will give
a theorem of purely geometric nature (valid in
{any theory of gravity} which assumes a Pseudo-Riemannian geometry as the
basis for the description of gravitational interactions)
that such wavy hair generated by the Garfinkle-Vachaspati
technique \cite{GV} cannot be detected by any scalar
invariant built out of the curvature and/or matter fields.
However in our example, we will show that
the tidal forces measured by an infalling observer diverge at
the `horizon' of the black string superposed with a vibration
in any mode with $l \ge 1$.
Hence the solutions with excited dipole or higher
multipole modes contain null singularities.
The same argument applied to the monopole mode shows that the leading order
tidal forces are finite. We will also construct a class of
monopole wave profiles for which the metric is manifestly smooth.
Hence at least some of the wavy strings indeed have regular horizons. 

The paper is organized as follows. In the next section, we
give a review of the wave generating technique, outlining
the conditions the matter distribution must satisfy for the
method to work. In section 3 we present our theorem on the elusive nature
of Garfinkle-Vachaspati waves, and give its detailed proof.
Section 4 contains the derivation
of the explicit form of the wave-black string solution and a brief discussion
of some of its properties. We show that these solutions contain
null singularities  for all higher multipole modes ($l\ge1$)
later in section 4.
In the last section, we construct an explicit
family of longitudinal wave ($l=0$) solutions for which
the metric which is analytic
on the horizon (and not only continuous, as shown in \cite{HM}).
Finally, we present our conclusions and consider directions for future
investigations.

\section{Garfinkle-Vachaspati Waves} \label{wavy}

In this section, we review the wave-generating technique of Garfinkle
and Vachaspati \cite{GV}. In doing so, we outline the situations in
which it may be applied, \ie the restrictions which must be satisfied by
the metric and matter fields. The crucial requirement is
that the solution possess a null, hypersurface-orthogonal, Killing
vector. The resulting presence of a null coordinate allows
one to effectively `linearize' Einstein's equations, and restore
hidden wave degrees of freedom.
The method was originally proposed for the Yang-Mills-Higgs
system coupled to gravity, and applied to straight cosmic
string solutions in four dimensions \cite{GV}. It was later extended
to gravity coupled to a scalar and a two-form potential in five
dimensions in the context of low energy string theory \cite{david}.
It was also used to construct similar wavy axionic strings in
refs.~\cite{DGHW,three}. The following description
will consider gravity in arbitrary dimensions coupled to
a matter sector including various scalars as well as a set of
different $p$-form potentials. Thus this discussion shows that this
technique is quite generally applicable in supergravity or low-energy
string theories \cite{sugra}.

Let us assume an action of the following form:
\beq\label{actor}
I=\int d^D\!x\sqrt{-g}
\left(R(g)-{1\over2}\sum_ah_a(\phi)(\nabla\phi_a)^2-{1\over2}\sum_p
f_p(\phi)F_{(p+1)}^2\right)\ .
\eeq
Thus as well as the metric, we have included a collection of scalar
fields $\phi_a$, which appear with non-derivative couplings in the
coefficient functions, $h_a(\phi)$ and $f_p(\phi)$. The above action
also involves a set of $p$-form potentials, $A_{(p)}$, through
their field strengths: $F_{(p+1)}=dA_{(p)}$. Hence there is a(n
Abelian) gauge invariance associated with these fields, $\delta
A_{(p)}=d\l_{(p-1)}$. Note that this action
is written in terms of the Einstein-frame metric, by which we mean
that there are no couplings to any $\phi_a$ appearing in the Ricci
scalar term of eq.~(\ref{actor}).
Now the gravity equations of motion may be written as
\beq\label{einst}
R^{\mu}{}_{\nu} - \frac{1}{2}\delta^{\mu}{}_{\nu} R =
{1\over2} T^{\mu}{}_{\nu}
\eeq
where
\beqa\label{stress}
T^{\mu}{}_{\nu} &=&\sum_a h_a(\phi)\left(g^{\mu\rho}\prt_\rho\phi_a
\prt_\nu\phi_a
-{1\over2}\delta^\mu{}_\nu\, (g^{\rho\l}\prt_\rho\phi_a\prt_\l\phi_a)\right)
\\
&+&\sum_p
 f_{p}(\phi)\left((p+1)\F{p+1}^{\mu\rho_1 \cdots \rho_p}
\F{p+1}_{\nu\rho_1 ... \rho_p} -
\frac{1}{2}\delta^{\mu}{}_{\nu}  \F{p+1}^{\rho_1 \cdots \rho_{p+1}}
\F{p+1}_{\rho_1 \cdots \rho_{p+1}}\right)
\nonumber
\eeqa
while the matter field equations may be written as
\beqa
\label{scalar} 0&=&
\prt_\mu \bigl(\sqrt{g}h_a(\phi)g^{\mu\nu}\prt_\nu\phi_a\bigr)
-{1\over2}\sqrt{g}\sum_b \frac{\partial h_{p}(\phi)}
{\partial \phi_a}\nabla^\mu\phi_b\nabla_\mu\phi_a\\
&&\qquad \quad \qquad - {1\over2}\sqrt{g}\sum_p \frac{\partial f_{p}(\phi)}
{\partial \phi_a}\F{p+1}^{\rho_1 \cdots \rho_{p+1}}
\F{p+1}_{\rho_1 \cdots \rho_{p+1}}\nonumber\\
\label{form} 0&=&
\prt_{\mu} \bigl(\sqrt{g}
f_{p}(\phi) \F{p+1}^{\mu\rho_1 \cdots \rho_p}\bigr)\ \ .
\eeqa
For later purposes, we have been explicit about the appearance
of the spacetime indices in these equations.

Now within this theory, let us assume a solution $(g,\phi_a,A_{(p)})$
for which there exists a vector field $k^\mu$ which is:
\beqa
\label{null} {\rm null:}&& k^\mu k_\mu=0\ ,\\
\label{hyper} {\rm hypersurface\ orthogonal:}
&&\nabla_{[\mu} k_{\nu]} = k_{[\mu}\nabla_{\nu]}\AH\ ,\\
\label{kill} {\rm and}\qquad{\rm Killing:}&&
\nabla_{(\mu} k_{\nu)} 
= 0\ .
\eeqa
Combining these equations, it is easy to show that
$k$ has a vanishing Lie-derivative on $\AH$,
\ie ${\cal L}_k \AH = k^{\mu} \prt_\mu \AH =0$.
Since we wish $k$ to yield an invariance of the full solution,
it is also assumed to have a vanishing Lie-derivative on
the matter fields,
\ie the matter fields are form-invariant along the flow of $k$.
Hence,
\beqa
\label{constr1}
{\cal L}_k \phi_a&=&k^\mu\prt_\mu\phi_a =0
\\
\label{constr2}
{\cal L}_k F_{(p+1)} &=& (d\,i_k+i_kd)F_{(p+1)}=d\,i_kF_{(p+1)}= 0
\eeqa
where the latter uses ${\cal L}_v=d\,i_v+i_vd$ on forms with
$i_v$ denoting the interior product, and the Bianchi identities,
$dF_{(p+1)}=0$. This form-invariance of the matter fields
guarantees that the stress-energy tensor
is form-invariant, as it must be given that
Einstein's equations (\ref{einst}) are satisfied. Note that
for the form fields, the vanishing Lie derivative is imposed
on the physical field strengths rather than the gauge-variant
potentials --- the latter may vary by a gauge transformation
along the $k$-flow, ${\cal L}_kA_{(p)}=d\l_{(p-1)}$.
In the following, we will further require that
these fields satisfy an additional transversality constraint,
namely,
\beq\label{transverse}
i_kF_{(p+1)}=k\wedge \theta_{(p-1)}\ .
\eeq
Here, the right-hand-side is the wedge product of one-form
$k_\mu dx^\mu$ with some ($p\,$--1)-form $\theta_{(p-1)}$, which necessarily
satisfies $i_k\theta_{(p-1)}=0$ since $(i_k)^2F_{(p+1)}=0$
--- see section 3.1, for details.

The solutions satisfying the above conditions can be interpreted
as gravity waves. Consider the coordinate system adapted to
the flow of $k$ --- as well as the cyclic coordinate $v$,
there is a coordinate $u$, given `roughly' by the integral of the
dual one-form $k=k_\mu dx^\mu$. The vanishing
Lie derivatives, eqs.~(\ref{kill}), (\ref{constr1})
and (\ref{constr2}),
simply means that none of the fields depend on $v$.
Hence the only `time' dependence can
arise through the null coordinate $u$,
and hence represents perturbations moving
at the speed of light in a certain direction of the space-time.
The Garfinkle-Vachaspati (GV) solution-generating technique
extends the solution by restoring
additional wave degrees of freedom. To do this,
they define a new metric by \cite{GV}
\begin{equation} \label{w31}
{g}'_{\mu\nu} = {g}_{\mu\nu} + \e^\AH\, \Psi\, k_{\mu}\,k_{\nu},
\end{equation}
while leaving all of the matter
fields unchanged\footnote{Throughout this
section, $k_\mu = g_{\mu\nu} k^\nu$. However, from the
definition of the metric $g'_{\mu\nu}$ and the fact that $k$ is
null, it doesn't matter which metric we use to lower and raise
the index of $k$.}.
The configuration $(g', \phi_a, A_{p})$ also yields a
solution provided 
the function $\Psi$ satisfies appropriate constraints. 
These restrictions will guarantee that after the metric is modified
the form of the explicit fields appearing in the equations
of motion remains unchanged. Hence eq.~(\ref{w31}) is promoted
to a map on the space of solutions. In order to
maintain the wave interpretation of the new solution, first one
requires that $k$ has a vanishing Lie-derivative on $\Psi$, \ie
$k^{\mu}\prt_{\mu}\Psi = 0$. One may verify that this ensures that
the hypersurface orthogonal and Killing conditions, eqs.~(\ref{hyper})
and (\ref{kill}), are still satisfied with the new metric
(with the same $\AH$). It is also obvious
that the null condition (\ref{null}) still holds with $g'$.
To determine what other restrictions must be imposed on
$\Psi$, one must consider the changes which the
map (\ref{w31}) induces in the equations of motion, 
eqs.~(\ref{einst}--\ref{form}). 

We begin by demonstrating that the matter field equations
are invariant under (\ref{w31}).
First note that the determinant of the metric would only be modified
by terms proportional to $k^{\mu}k_{\mu}$. However since $k$ is null, 
the determinant is invariant: $\det({g}')=\det({g})$.
On the other hand, the inverse to the metric $g'_{\mu\nu}$ is
given by
\begin{equation} \label{inversem}
g'^{\mu\nu} = g^{\mu\nu} - \e^\AH\, \Psi\, k^{\mu}k^{\nu}\ .
\end{equation}
Now given the transversality constraint (\ref{transverse}), we see
that raising all of the indices of the form fields with the new
metric yields
\beqa\label{raise}
\Fp{p+1}^{\mu_1\cdots\mu_{p+1}}&=&\F{p+1}^{\mu_1\cdots\mu_{p+1}}
+p(p+1)\,k^{[\mu_1}k^{\mu_2}\T{p-1}^{\mu_3\cdots\mu_{p+1}]}
\nonumber\\
&=&\F{p+1}^{\mu_1\cdots\mu_{p+1}}
\eeqa
where in the first line, terms with more powers of $k$ have automatically
vanished since $i_k\theta_{(p-1)}=0$. Given this result as well
as the invariance of det($g$), it is clear that the equations of motion
for the form fields (\ref{form}) are left unchanged by the map (\ref{w31}).
Further given the vanishing Lie derivative (\ref{constr1}), it is clear that
$g'^{\mu\nu}\prt_\nu\phi_a=g^{\mu\nu}\prt_\nu\phi_a$ and hence the
scalar equations (\ref{scalar}) are also unchanged. Thus we have shown that
the configuration $(g', \phi_a, A_{p})$ provides a solution of all
of the matter field equations. 

The same considerations as above also show that the stress-energy
tensor with mixed indices $T^{\mu}{}_{\nu}$, given in eq.~(\ref{stress}),
remains unchanged when the metric is modified as in eq.~(\ref{w31}).
The only step which remains in examining the gravity equation
is to compute the change of the mixed-index
Einstein tensor. This is reduced to computing the change
in the Ricci tensor only, because the Ricci scalar may be
eliminated from eq.~(\ref{einst})
using $R = \frac{1}{2-D}T^{\mu}{}_{\mu}$ in $D$ dimensions.
It is straightforward to calculate that the Christofel symbols
for the two metrics are 
related by ${\Gamma'}^{\mu}_{\nu\lambda} =
{\Gamma}^{\mu}_{\nu\lambda} + {\Omega}^{\mu}_{\nu\lambda}$
where
\begin{equation} \label{omega}
{\Omega}^{\mu}_{\nu\lambda} = \frac{1}{2} \Bigl(
\nabla_{\nu}\bigl(\e^\AH \Psi k^{\mu} k_{\lambda}\bigr)
+\nabla_{\lambda}\bigl(\e^\AH \Psi k^{\mu} k_{\nu}\bigr)
-\nabla_{\mu}\bigl(\e^\AH \Psi k^{\nu} k_{\lambda}\bigr)
\Bigr)\ .
\end{equation}
Further using the properties of $k$, as well as $k^\mu\prt_\mu \AH=0
=k^\mu\prt_\mu\Psi$, one can show that
$\Omega^{\mu}_{\mu\nu} =0= k^\mu \nabla_\lambda
\Omega^{\lambda}_{\mu\nu} = \Omega^{\rho}_{\mu\lambda}
\Omega^{\lambda}_{\rho\nu}$. Thus
one finds that ${R'}_{\lambda\nu} =
{R}_{\lambda\nu} + \nabla_\rho \Omega^{\rho}_{\lambda\nu}$
and hence
\begin{equation} \label{mixedr}
{R'}^{\mu}{}_{\nu} = {R}^{\mu}{}_{\nu} +
\e^\AH \Psi k^{\mu} k^{\lambda}{R}_{\lambda\nu} +
\nabla_\rho \bigl( \Omega^{\rho}_{\lambda\nu} g^{\mu\lambda} \bigr)
\end{equation}
Again using the properties of $k$, one can show:
$k^\lambda R_{\lambda\nu} = k_\nu \nabla^2 \AH /2$. Substituting this
into eq.~(\ref{mixedr}) and after a few more manipulations, we finally
arrive at \cite{GV}:
\begin{equation} \label{mixedr1}
{R'}^{\mu}{}_{\nu} = {R}^{\mu}{}_{\nu} - \frac{1}{2}
\e^\AH k^{\mu} k_{\nu}\nabla^2 \Psi
\end{equation}
Therefore, the variation of the mixed Ricci tensor
${R}^{\mu}{}_{\nu}$ under (\ref{w31}) is
proportional to $\nabla^2 \Psi$, and so vanishes if we demand
that $\Psi$ solves the covariant Laplace equation (in the
background defined by $g_{\mu\nu}$).

Therefore given a solution for which eqs.~(\ref{null}--\ref{transverse}) are
satisfied, then eq.~(\ref{w31}) provides a map to a new solution
provided
\beq\label{psiconstr}
k^\mu\prt_\mu\Psi=0 \qquad{\rm and}
\qquad \nabla^2 \Psi=0\ \ .
\eeq
We note that the condition $\nabla^2 \Psi=0$ is really just the
spatial Laplace equation since
$k^\mu \prt_\mu \Psi =0$. Furthermore the latter indicates that
the moduli of the new solutions only depend the retarded
time -- $u$, above -- and therefore they still represent 
gravitational waves.

The preceding discussion was phrased in terms of the Einstein-frame
metric. However, in some instances (such as the example in section
\ref{solex}), it is more convenient to work in terms of a conformally
related metric, \eg
\beq\label{confu}
\tilde{g}_{\mu\nu}=\e^{\al(\phi)}g_{\mu\nu}\ \ .
\eeq
Given in terms of $\tilde{g}$, the action will contain non-minimal
couplings of the scalar fields to the Ricci scalar. It is straightforward
to show that most of the constraints (\ref{null}--\ref{transverse})
are unchanged when written in terms of the conformally transformed
metric. The only
change is that the hypersurface orthogonal condition (\ref{hyper}) becomes
\beq\label{hypercon}
\tilde{\nabla}_{[\mu} \tilde{k}_{\nu]} = 
\tilde{k}_{[\mu}\nabla_{\nu]}\tilde{\AH}
\qquad{\rm where}\ \ \tilde{\AH}=\AH-\al(\phi)\ \ .
\eeq
Here we have denoted $\tilde{k}_\mu=\tilde{g}_{\mu\nu}k^\nu$.
Thus the map (\ref{w31}) becomes
\beq\label{w311}
{\tilde{g}}'_{\mu\nu} = \tilde{g}_{\mu\nu} + \e^{\tilde{\AH}}\,
\Psi\, \tilde{k}_{\mu}\,\tilde{k}_{\nu},
\eeq
which is equivalent to ${\tilde{g}}'_{\mu\nu} =\e^{\al(\phi)}g'_{\mu\nu}$.
Finally, of course, the constraints on $\Psi$, which ensure that
eq.~(\ref{w311}) provides a solution of the gravitational equations of
motion are identical to those appearing in eq.~(\ref{psiconstr}). The
latter may be written as follows, when expressed 
in terms of $\tilde{g}$:
\beq\label{psicon2}
k^\mu\prt_\mu\Psi=0 \qquad{\rm and}
\qquad \prt_\mu\left(\e^{{2-D\over2}\al(\phi)}\sqrt{-\tilde{g}}
{\tilde{g}}^{\mu\nu}\prt_\nu \Psi\right)=0
\eeq
where again $D$ denotes the dimension of the spacetime.

Before concluding this section, let us mention a few simple extensions
of the original action (\ref{actor}) for which the preceding discussion
would still be applicable. First, we could add a scalar potential
$U(\phi)$ (which could then include a cosmological constant) to
eq.~(\ref{actor}). This would modify the stress-energy (\ref{stress})
by a term proportional to $\delta^\mu{}_\nu U(\phi)$, as well as adding
a $\prt_{\phi_a}U$ term to the scalar equations (\ref{scalar}).
Both of these terms are obviously invariant under the map (\ref{w31})
and so the above construction is still valid.

Another non-minimal coupling, which commonly arises among
the form fields, are Chern-Simons-like terms appearing in the
definition of the field strengths, \eg for some choice of
$m$ and $n$, $F_{(m+n+1)} = dA_{(m+n)} +  \al A_{(m)} \wedge  dA_{(n)}$
where we still assume $F_{(m+1)}=dA_{(m)}$ and $F_{(n+1)}=dA_{(n)}$.
This definition leads to two modifications in the equations of motion.
(Note that for $F_{(m+n+1)}$ the Bianchi identity is also modified,
$dF_{(m+n+1)}=\al F_{(m+1)}\wedge F_{(n+1)}$, but this has
no consequences for the present discussion.)
First, eq.~(\ref{form}) is modified for $p=m$ by the
introduction of a source term which is proportional to
\beq\label{source}
\sqrt{g}\,\F{m+n+1}^{\rho_1\cdots\rho_m\mu_1\cdots\mu_{n+1}}
\F{n+1}_{\mu_1\cdots\mu_{n+1}}\ \ .
\eeq
With 
the same transversality constraint on $F_{(m+n+1)}$ as above,
eq.~(\ref{transverse}), we still conclude that its form with
all indices raised remains invariant, as in eq.~(\ref{raise}), and
hence this new term is also unchanged by eq.~(\ref{w31}). The second change
is the appearance of a source term in eq.~(\ref{form}) for $p=n$.
In this case upon applying the equation of motion (\ref{form}) for
$p=n+m$, this second source term takes precisely the same form
as above, merely interchanging the roles of $m$ and $n$. Therefore
it is also invariant under the map (\ref{w31}). Thus the
GV procedure will still be valid for solutions
where these Chern-Simons-like couplings make nontrivial contributions.

The final extension which we will consider is the addition of 
topological interactions, \eg
\[
\int A_{(\ell)}\,dA_{(m)}\,dA_{(n)}
\]
where $\ell+m+n+2=D$. Such an interaction again modifies the 
equations of motion (\ref{form}) for $p=\ell,m,n$ by the introduction
of source terms of the form, \eg
\[
\sqrt{g}\,
\pol^{\mu_1\cdots\mu_\ell\nu_1\cdots\nu_{m+1}\rho_1\cdots\rho_{n+1}}
\F{m+1}_{\nu_1\cdots\nu_{m+1}}\F{n+1}_{\rho_1\cdots\rho_{n+1}}
\]
where $\pol$ is the Levi-Civita tensor in $D$ dimensions. Given the
invariance of the determinant of the metric, $\pol'=\pol$ as forms.
Hence we must compare raising all of the indices  with
$g'$ and $g$. Given eq.~(\ref{inversem}), we see
\beqa\label{raisevol}
\pol'^{\mu\nu\cdots\rho}&=&\pol^{\mu\nu\cdots\rho}
-D\,e^\AH\Psi\,k^{[\mu|}k_\l\pol^{\l|\nu\cdots\rho]}
\nonumber \\
&=&\pol^{\mu\nu\cdots\rho}
\eeqa
where the vanishing of the second term in the first line relies
on a result proven in the following section --- see eq.~(\ref{levee}).
Therefore the new source terms are
again unchanged by the map (\ref{w31}), and so these topological
interactions provide no obstruction for this solution generating
technique.

\section{The Elusive Wave} \label{long}

Generically it is the case that the original solution and
that carrying a wave induced by eq.~(\ref{w31}) are not diffeomorphic,
however, the wave turns out to be very elusive. In this section, we will
present a theorem of a very general nature, showing that all
the scalar curvature invariants of
the two metrics related by the wave-generating technique are
in fact identical. Thus, no curvature invariant
of the oscillating metric can  be used to detect the presence of
the gravitational wave! The theorem generalizes in a straightforward
way to include any scalar invariants constructed from
both the metric and matter fields.

To prove this result, we need only assume the existence of a
metric $g_{\mu\nu}$ which admits a null, hypersurface-orthogonal,
Killing vector. We do not require that the metric solves Einstein's
or any other equations. Thus our result is purely geometrical
in nature, and holds for any metric satisfying the symmetry condition.
The precise statement of the theorem is as follows:

{\it If  $g_{\mu\nu}$ is a pseudo-Riemannian metric admitting
a null, hypersurface-orthogonal, Killing vector $k^\mu$, and
$g'_{\mu\nu}= g_{\mu\nu} + \kappa\, k_{\mu}k_{\nu}$ where
$\kappa$ is any scalar Lie-derived by $k$ to zero, \ie
${\cal L}_k \kappa = 0$, then all of the scalar curvature invariants of
$g'_{\mu\nu}$ are exactly identical to the corresponding
curvature invariants of $g_{\mu\nu}$. }

Hence the GV transform
(\ref{w31}) provides one example of such metrics with $\kappa = \e^\AH \Psi$.
In the following, we will refer to $g_{\mu\nu}$ as the original
metric, $g'_{\mu\nu}$ as simply the primed or shifted metric,
and $\kappa$ as the wave profile, in analogy to the last section.

Before presenting the details, let us
first give a brief sketch of our proof.
Before examining the curvature invariants, we establish two crucial
results. First, contracting $k^\mu$ with any tensor
constructed from the original metric, its curvature and covariant
derivatives of the curvature and/or any scalars with a vanishing
Lie-derivative under $k$, produces a sum of terms in each of which $k$
appears uncontracted. Second,
any tensor (\eg the curvature or covariant derivatives 
thereof) in the primed background may be written
as the sum of that for the original metric plus a $\kappa$-dependent
term, for which all of the contributions are at least bilinear in
the Killing vector $k$. These two results set the stage for an examination
of the scalar curvature invariants. There we find that all of the
new $\kappa$-dependent terms vanish by combining the previous two
results with the fact that $k$ is null.
Hence, we conclude  that the original and
the corresponding primed invariants are identical.
As a consequence, no evidence of 
the wave profile $\kappa$ can be detected in 
any of the curvature invariants of the metric. 
This result represents a generalization of the previous studies 
of curvature invariants of geometries admitting null,
{covariantly constant,} Killing vectors \cite{SH,S}.
An immediate corollary of this theorem is that if $g_{\mu\nu}$ represents
an extended black object with a regular horizon to which we
add a GV wave as in the last section, then the oscillations
will not produce new {\it scalar} curvature singularities --- however, in
later sections, we will
discuss the limitations of this statement with explicit examples.

\medskip
\noindent{\it 0) Useful Formulae}

Let us start by listing some important formulae which arise
from the existence of a null, hypersurface-orthogonal, Killing vector.
As stated before (in eq.~(\ref{hyper})), the hypersurface-orthogonal
condition
amounts to $\nabla_{[\mu} k_{\nu]} = k_{[\mu} \nabla_{\nu]} \AH$ for some
scalar $\AH$ which can be determined from the metric. Combining
this with the Killing condition (\ref{kill}) yields
\begin{equation} \label{w30a}
\nabla_{\mu} k_{\nu} = \frac{1}{2}\bigl(k_{\mu}\nabla_{\nu}\AH
- k_{\nu}\nabla_{\mu}\AH\bigr)\ .
\end{equation}
Further, it is not
difficult to see that the Killing condition alone leads to
\beq\label{curvy}
\nabla_\nu \nabla_\mu k_\lambda = k^\rho R_{\rho\nu\mu\lambda}
\eeq
as can be determined by considering the
commutator of two covariant derivatives acting on $k$.
We also note that we may express
\beqa\label{interest}
k^\rho R_{\rho\nu\lambda\sigma} &=& \nabla_\sigma \nabla_\lambda k_\nu -
\nabla_\lambda \nabla_\sigma k_\nu \nonumber\\
&=& k_{[\lambda} \nabla_{\sigma]}\nabla_\nu \AH -{1\over2} 
k_{[\lambda} \nabla_{\sigma]} \AH\nabla_\nu \AH
\eeqa
using eq.~(\ref{w30a}).
Recall the expression for the Lie-derivative
of a general tensor $T^{\mu_1 \ldots \mu_p}{}_{\nu_1 \ldots \nu_q}$:
\begin{eqnarray}
\label{Lieder}
{\cal L}_v T^{\mu_1 \ldots \mu_p}{}_{\nu_1 \ldots \nu_q} &=& v^\lambda
\nabla_\lambda
T^{\mu_1 \ldots \mu_p}{}_{\nu_1 \ldots \nu_q} \nonumber \\
&&- T^{\lambda  \ldots \mu_p}{}_{\nu_1 \ldots \nu_q} \nabla_\lambda v^{\mu_1} 
- \ldots -
T^{\mu_1 \ldots \lambda}{}_{\nu_1 \ldots \nu_q} \nabla_\lambda v^{\mu_p}\\
&&+ T^{\mu_1 \ldots \mu_p}{}_{\lambda \ldots \nu_q} \nabla_{\nu_1} v^{\lambda}
+ \ldots
+ T^{\mu_1 \ldots \mu_p}{}_{\nu_1 \ldots \lambda} \nabla_{\nu_q}
v^{\lambda}\ \ .
\nonumber
\end{eqnarray}
From this definition and the Killing condition (\ref{kill}), one may
show that the Lie-derivative with respect
to a Killing vector ${\cal L}_k$ commutes with the covariant
derivative. Begin by considering:
\begin{eqnarray}\label{commcalc}
\nabla_\mu {\cal L}_k T_{\nu_1 \ldots \nu_q} &=& k^\lambda \nabla_\mu
\nabla_\lambda
 T_{\nu_1 \ldots \nu_q} + \nabla_\lambda  T_{\nu_1 \ldots \nu_q}
\nabla_{\mu} k^{\lambda} + \ldots
+ \nabla_\mu  T_{\nu_1 \ldots \lambda} \nabla_{\nu_q} k^{\lambda} \nonumber \\
& &\ + T_{\lambda \ldots \nu_q} \nabla_\mu \nabla_{\nu_1} k^{\lambda} + \ldots
+ \nabla_\mu  T_{\nu_1 \ldots \lambda} \nabla_\mu
\nabla_{\nu_q} k^{\lambda} \nonumber \\
&=& k^\lambda  \nabla_\lambda\nabla_\mu 
 T_{\nu_1 \ldots \nu_q} + \nabla_\lambda  T_{\nu_1 \ldots \nu_q}
\nabla_{\mu} k^{\lambda} + \ldots
+ \nabla_\mu  T_{\nu_1 \ldots \lambda} \nabla_{\nu_q} k^{\lambda} \nonumber \\
& &\ + \Bigl(T_{\lambda \ldots \nu_q} (R^{\lambda}{}_{\nu_1\rho\mu} +
R_{\rho\mu\nu_1}{}^{\lambda}) k^{\rho} + \ldots
+ T_{\nu_1 \ldots \lambda} (R^{\lambda}{}_{\nu_q\rho\mu} +
R_{\rho\mu \nu_q}{}^{\lambda}) k^{\rho} \Bigr) \nonumber \\
&=& {\cal L}_k \nabla_\mu T_{\nu_1 \ldots \nu_q} 
\end{eqnarray}
using eq.~(\ref{curvy}) and the standard commutator:
$[\nabla_{\mu},\nabla_{\lambda}] T_{\nu_1 \ldots \nu_q}=
R^{\rho}{}_{\nu_1\lambda\mu}  T_{\rho \ldots \nu_q} + \ldots\ $.
Hence, $[{\cal L}_k, \nabla_\mu] T_{\nu_1 \ldots \nu_q}=0$.
The case where $[{\cal L}_k, \nabla_\mu]$ acts on a tensor with some
raised indices is trivially
related to this one because ${\cal L}_k g^{\nu\lambda}=0=
\nabla_\mu g^{\nu\lambda}$.
Similarly, one can show ${\cal L}_k R_{\rho\lambda\nu\mu} =0$, 
as well as ${\cal L}_k \AH = k^\mu\nabla_\mu\AH=0$. 
Further recall by definition,
${\cal L}_k \kappa =k^\mu\nabla_\mu\kappa=0$.
We will find all of these formulae useful below,
when evaluating the contractions of tensors with the null vector $k$.

\medskip
\noindent{\it 1) Contraction Identities}

The first step is to show that for any tensor built out of the original
curvature, any scalars with vanishing Lie-derivative
under $k$ (\ie $\AH$ and $\kappa$), and
any number of covariant derivatives (with respect to the
original connection) acting on either of these,
the contraction $k^{\mu} T_{\nu_1 \ldots \nu_p \mu
\lambda_1 \ldots \lambda_q}$
is at least linear in vector $k_{\mu}$, \ie it can be expressed as a
linear combination of terms which factorize as some
tensor of rank lower by two and the vector $k$:
\begin{equation}\label{factor}
k^{\mu} T_{\nu_1 \ldots \nu_p \mu \lambda_1 \ldots \lambda_q} =
\sum^{p}_{n=1} k_{\nu_n} \thetn{n}_{\nu_1 \ldots \underline{\nu_n} \ldots
\nu_p \lambda_1 \ldots \lambda_q}
+ \sum^{q}_{n=1} k_{\lambda_n} \thetn{p+n}_{\nu_1 \ldots \nu_p \lambda_1
\ldots\underline{\lambda_n} \ldots \lambda_q}
\end{equation}
where underlining an index denotes its deletion from the expression.
(For convenience, we will work with only covariant, \ie `downstairs',
indices.)
This is true in the few simple cases encountered so far, \eg
$k^{\mu} \nabla_{\mu} \AH =0=k^\mu\nabla_\mu
\kappa$ and $k^\rho R_{\rho\nu\lambda\sigma}$ as given in eq.~(\ref{interest}).
Let us now show that it holds in general for what we denote as
`primary' tensors, namely tensors obtained by an arbitrary
number of covariant derivatives acting on the curvature or the scalars $\AH,
\kappa$. The proof again relies on mathematical
induction, and makes essential use of eq.~(\ref{w30a}) which
allows any covariant derivative of $k$  to be re-expressed in terms
on undifferentiated $k$'s.
First, we establish the result for the simplest cases:
for any scalar $B$ Lie-derived to zero by $k$, we have
$k^{\mu} \nabla_{\mu} B =0$ and further
\beqa\label{onemore}
k^{\mu}\nabla_{\nu}\nabla_{\mu}B=
k^{\mu} \nabla_{\mu} \nabla_{\nu} B &=& \nabla_\nu\left(k^{\mu} \nabla_{\mu}B
\right)- \nabla_\mu B \nabla_\nu k^{\mu}
\nonumber\\
&=&- {1\over2} k_{\nu}\, \nabla^\mu \AH \nabla_\mu B
\eeqa
using eq.~(\ref{w30a}). Now combining ${\cal L}_k
R_{\alpha\beta\gamma\sigma}=0$ and eq.~(\ref{interest}),
we find
\begin{eqnarray}\label{riemconct1}
k^{\mu} \nabla_{\mu}  R_{\alpha\beta\gamma\sigma} &=&
R^{\mu}{}_{\beta\gamma\sigma}  k_{[\mu} \nabla_{\alpha]} \AH + \ldots +
R_{\alpha\beta\gamma}{}^{\mu} k_{[\mu} \nabla_{\sigma]}\AH  \nonumber \\
&=& k_{\alpha}\thetn{1}_{\beta\gamma\sigma} + \ldots + k_\sigma
\thetn{4}_{\alpha\beta\gamma}
\end{eqnarray}
where in the second line we collect the like terms. Explicitly, one finds, \eg
$\thetn{1}_{\beta\gamma\sigma}=-{1\over2}(\nabla^\mu R_{\mu\beta\gamma\sigma}
+\nabla_\beta\nabla_{[\gamma}S\,\nabla_{\sigma]}S)$ and
$\thetn{2}_{\al\gamma\sigma}=-{1\over2}(\nabla^\mu R_{\al\mu\gamma\sigma}
-\nabla_\al\nabla_{[\gamma}S\,\nabla_{\sigma]}S)$.
Similarly,
\begin{eqnarray}\label{riemconct2}
k^{\alpha} \nabla_{\mu}  R_{\alpha\beta\gamma\sigma} &=&  \nabla_{\mu} (
k^{\alpha} R_{\alpha\beta\gamma\sigma}) -
R_{\alpha\beta\gamma\sigma} \nabla_{\mu} k^{\alpha} \nonumber \\
&=& k_\mu\bthetn{1}_{\beta\gamma\sigma} + \ldots +k_\sigma
\bthetn{4}_{\mu\beta\gamma}
\end{eqnarray}
again using eqs.~(\ref{w30a}) and (\ref{interest}).

So now let us assume that eq.~(\ref{factor}) holds for all primary
tensors of rank $q$ or less.
Now by our definition, a primary tensor
of rank $q+1$ will be obtained by covariant derivative acting
on a primary tensor of rank $q$, \ie 
$\nabla_{\lambda}T_{\lambda_1 \ldots \lambda_q}$.
Hence we consider
\begin{eqnarray}\label{induc}
k^\mu \nabla_{\lambda_1} T_{\mu\lambda_2 \ldots \lambda_q} &=&
\nabla_{\lambda_1}\left(k^\mu  T_{\mu\lambda_2 \ldots \lambda_q}\right)
- T_{\mu\lambda_2 \ldots \lambda_q}  \nabla_{\lambda_1} k^\mu  \nonumber \\
&=& \sum^{q}_{n=2}  \nabla_{\lambda_1} \left( k_{\lambda_n} 
\thetn{n}_{\lambda_2  \ldots\underline{\lambda_n} \ldots \lambda_q}\right)
+ \frac{1}{2}  T_{\mu\lambda_2 \ldots \lambda_q}  k^{\mu} \nabla_{\lambda_1}\AH
- \frac{1}{2}  T_{\mu\lambda_2 \ldots \lambda_q}  k_{\lambda_1} \nabla^\mu \AH
\nonumber \\
&=& \sum^{q}_{n=2}  \left(k_{\lambda_n}\nabla_{\lambda_1}\thetn{n}_{\lambda_2
\ldots\underline{\lambda_n} \ldots \lambda_q}
+ \nabla_{\lambda_1} k_{\lambda_n} \thetn{n}_{\lambda_2  \ldots
\underline{\lambda_n} \ldots \lambda_q} 
+ \frac{1}{2}  k_{\lambda_n}  \thetn{n}_{\lambda_2  \ldots
\underline{\lambda_n} \ldots \lambda_q}\nabla_{\lambda_1}\AH\right)
\nonumber \\
&& \qquad- \frac{1}{2} k_{\lambda_1} T_{\mu\lambda_2 \ldots \lambda_q} 
\nabla^\mu \AH
\\
&=& \sum^{q}_{n=2} \left( k_{\lambda_n} \nabla_{\lambda_1}\thetn{n}_{\lambda_2
\ldots\underline{\lambda_n} \ldots \lambda_q}
+ \frac{1}{2} k_{\lambda_1}\nabla_{\lambda_n}\AH\,\thetn{n}_{\lambda_2  \ldots
\underline{\lambda_n} \ldots \lambda_q}\right)
- \frac{1}{2} k_{\lambda_1} T_{\mu\lambda_2 \ldots \lambda_q}
\nabla^\mu \AH
\nonumber \\
&=& \sum^{q}_{n=1}  k_{\lambda_n} 
\bthetn{n}_{\lambda_1  \ldots \underline{\lambda_n} \ldots
\lambda_q} \nonumber
\end{eqnarray}
after collecting the like terms into the tensors $\bthetn{n}$.
Of course, the same result follows if $k^\mu$ is contracted
with any of the other indices on $T$ above.
The last possibility is $k^{\mu} \nabla_{\mu} T_{ \lambda_1 \ldots
\lambda_q}$. Here we use ${\cal L}_k T_{\lambda_1 \ldots \lambda_q} = 0$
which holds by $[{\cal L}_k, \nabla_{\mu}]=0$ and the fact that
both $S$ and the curvature have a vanishing Lie-derivative under $k$.
Thus,
\begin{eqnarray}\label{induc2}
k^\mu \nabla_{\mu} T_{\lambda_1 \ldots \lambda_q} &=& - T_{\mu\lambda_2
\ldots \lambda_q} \nabla_{\lambda_1} k^\mu -
\ldots - T_{\lambda_1 \ldots \lambda_{q-1}\mu}  \nabla_{\lambda_q} k^\mu
 \nonumber \\
&=& \frac{1}{2}\left( T_{\mu\lambda_2 \ldots \lambda_q} \nabla_{\lambda_1}\AH+
\ldots + T_{\lambda_1 \ldots \lambda_{q-1}\mu}  \nabla_{\lambda_q} \AH\right)
k^\mu \\
&& \quad-  \frac{1}{2}\left( T_{\mu\lambda_2 \ldots \lambda_q}k_{\lambda_1}+
\ldots + T_{\lambda_1 \ldots \lambda_{q-1}\mu}  k_{\lambda_q}\right)
\nabla^\mu \AH 
\nonumber\\
&=&\sum^{q}_{n=1} k_{\lambda_n}  \tilde{\theta}^{(n)}_{\lambda_1 \ldots
\underline{\lambda_n} \ldots \lambda_{q}} \nonumber
\end{eqnarray}
where we applied eq.~(\ref{factor}) for rank $q$ primary tensors in the
second line and gathered the like terms. With eqs.~(\ref{induc}) and
(\ref{induc2}), we have shown that
eq.~(\ref{factor}) holds for any index $\mu$ on a primary tensor of rank $q+1$.
Therefore by induction this factorization property is established for
all primary tensors.

At this stage, we must consider the `secondary' tensors, \ie
the $\theta$ tensors produced in the
contractions of $k$ with primary tensors --- although here we will leave many
of the details to the reader. Considering the simplest examples in
eqs.~(\ref{interest}) and (\ref{onemore}--\ref{riemconct2}), one finds that
the $\theta$ tensors are not simply primary tensors, but rather involve
certain products and/or contractions of primary tensors. However,
in those particular examples, it is not hard to show that they
share two important properties in common with the primary
tensors: {\it i)}
the $\theta$'s have a vanishing Lie-derivative under $k$, and
{\it ii)} upon contraction with $k$, they factorize as in eq.~(\ref{factor}).
Having established that these conditions
apply in the simplest cases, it is straightforward to
formulate an inductive proof to show that they also apply
for the $\theta$ tensors produced from primary tensors of
higher rank. One would begin by assuming that {\it (i)} and {\it (ii)}
hold for the $\theta$'s arising from rank $q$ primary tensors, and
then examine those produced at rank $q+1$. In the case considered in
eq.~(\ref{induc}), one has
\beqa\label{thetan}
\bthetn{1}_{\l_2\ldots\l_q}&=&
{1\over2}\sum_{n=1}^q \nabla_{\l_n}S\,\thetn{n}_{\l_2\ldots
\underline{\l_n}\ldots\l_q}-{1\over2}\nabla^\mu S\,T_{\mu\l_2
\ldots\l_q}\nonumber\\
\bthetn{n}_{\l_1\ldots\underline{\l_n}\ldots\l_q}&=&
\nabla_{\l_1}\thetn{n}_{\l_2\ldots\ldots\l_q}
\qquad\qquad{\rm for\ }n>1\ .
\eeqa
It is easy to show that $\bthetn{1}$ will satisfy both the Lie-derivative
and factorization properties because all of the components of which
it is comprised (\ie $\theta^{(n)}$'s, $T$, $\nabla S$) do.
For $n>1$, $\bthetn{n}$ will be Lie-derived to zero by $k$ because
$\thetn{n}$'s are and $[{\cal L}_k, \nabla_{\mu}]=0$. Given the vanishing
Lie-derivative and that factorization (\ref{factor}) holds for the
$\theta^{(n)}$'s, one would extend this condition to the $\bthetn{n}$
tensors in the same way as in eqs.~(\ref{induc}) and (\ref{induc2})
for the primary tensors. One should then examine the $\tilde\theta$'s
arising in eq.~(\ref{induc2}) for that particular case of the rank $(q+1)$
primary tensors. One is again able to show that both {\it (i)} and {\it (ii)}
applies for these $\theta$ tensors as well, although the index gymnastics
is somewhat more involved. Furthermore, the proof of these properties extends
in a similar way to any `higher-order' tensors, that is any new
$\theta$'s produced by contracting $k$ with $\theta$ tensors.

Now this intermediate result for the $\theta$ tensors is necessary
in order to show that the factorization property (\ref{factor}) also
holds for
products of primary tensors, with arbitrary contractions of pairs
of indices. Our principal tool here is again mathematical induction.
We begin by considering quantities of the simple product
$T^{(1)}_{\nu_1\ldots\nu_{p+1}\lambda_1\ldots\lambda_q}
T^{(2)}{}^{\lambda_1\ldots\lambda_q}{}_{\omega_1\ldots\omega_l}$,
where both $T^{(1)}$ and $T^{(2)}$ are primary tensors.
Using the properties of primary tensors, we find
\begin{eqnarray}\label{secon}
k^{\mu} T^{(1)}_{\nu_1\ldots\mu\ldots\nu_p\lambda_1\ldots\lambda_q}
T^{(2)}{}^{\lambda_1\ldots\lambda_q}{}_{\omega_1\ldots\omega_l} &=&
 \sum^{p}_{n=1} k_{\nu_n} \thetn{n,1}_{\nu_1\ldots \underline{\nu_n} \ldots
\nu_p\lambda_1\ldots\lambda_q}
T^{(2)}{}^{\lambda_1\ldots\lambda_q}{}_{\omega_1\ldots\omega_l}\\ 
&& + \sum^{q}_{n=1} k_{\lambda_n} \thetn{p+n,1}_{\nu_1\ldots
\nu_p\lambda_1\ldots\underline{\lambda_n} \ldots \lambda_q}
T^{(2)}{}^{\lambda_1\ldots\lambda_q}{}_{\omega_1\ldots\omega_l}  \nonumber  \\
\end{eqnarray}
Here the first term has the required form, and so
for $q=0$, \ie no contractions, we would have the desired result.
However, for $q\ge1$,
we need to look at the cross-terms in the second sum \eg
\begin{eqnarray}\label{secon2}
k_{\lambda_1}\thetn{p+1,1}_{\nu_1\ldots \nu_p \lambda_2\ldots \lambda_q} 
T^{(2)}{}^{\lambda_1\ldots\lambda_q}{}_{\omega_1\ldots\omega_l} &=&
\thetn{p+1,1}_{\nu_1\ldots \nu_p \lambda_2\ldots \lambda_q}
\sum^{q}_{n=2} k^{\lambda_n} \thetn{n,2}{}^{\lambda_2\ldots
\underline{\lambda_n}\ldots \lambda_q}{}_{\omega_1\ldots\omega_l}
\nonumber \\
&& + \thetn{p+1,1}_{\nu_1\ldots \nu_p \lambda_2\ldots \lambda_q}
\sum^{l}_{n=1} k_{\omega_n} \thetn{q+n,2}{}^{\lambda_1\ldots
\lambda_q}{}_{\omega_1\ldots \underline{\omega_n} \ldots\omega_l}
\end{eqnarray}
Here we will have arrived at the desired form if $q=1$, but
for $q\ge2$, we have generated further cross-terms in which $k$ 
is contracted back on $\thetn{p+1,1}$. However, since 
the $\theta$'s also factorize according to eq.~(\ref{factor})
as described above, we may continue this procedure. Now
given that $q$, the number of contractions, is finite, 
this `ladder' of $k$ contractions will eventually terminate, since at
each step the number of contracted index pairs is reduced by one in
each of the subsequent cross terms. 
Therefore after a finite number of steps, we arrive at a factorization
without any contractions yielding the desired form.
Given this result, it is obvious that we may in a similar way
extend eq.~(\ref{factor})
to apply for an arbitrary product of primary tensors, including
arbitrary contractions.

To conclude this subsection, we consider contractions of $k$ with the
Levi-Civita tensor, $\pol$, \ie the volume form on the $D$-dimensional
spacetime. To analyze this case, we resort to
local coordinate patches adapted to the properties of $k$. First, the
Killing condition (\ref{kill}) indicates that we can find a cyclic
coordinate such that $k^\mu\prt_\mu=(\prt/\prt v)$. Next the hypersurface
orthogonal condition (\ref{hyper}) indicates that we can find a dual
coordinate such that $k_\mu dx^\mu=\e^{-S}du$. In this local
coordinate patch, one of the free indices in $k^\mu\pol_{\mu\al\cdots\beta}$
must then take the value $u$, and hence we can write 
\beq\label{levee}
i_k\pol = k\wedge\theta
\eeq
where in this case $\theta$ is some ($D$--2)-form.
However, as written this result is coordinate independent
and so must hold in general.\footnote{Eq.~(\ref{levee}) is the
essential result required to prove eq.~(\ref{raisevol}).}
Thus we have shown that the Levi-Civita tensor factorizes in the
same way as the primary tensors.

In fact, the resulting $\theta$ satisfies
eq.~(\ref{factor}) as well in a trivial way since $i_k\theta=0$.
This result is again most easily derived using the local coordinates
introduced above. Because of the antisymmetry of the Levi-Civita
tensor, none of the indices on the right-hand-side of eq.~(\ref{levee})
correspond to the cyclic coordinate $v$. Hence $i_k\theta=0$ with this
choice of coordinates, but this equation must then be valid in general.

As an aside, we note that the preceding discussion is equally
applicable for the $\theta_{(p-1)}$ arising in the transversality
constraint (\ref{transverse}): $i_kF_{(p+1)}=k\wedge \theta_{(p-1)}$.
In this case using the adapted
coordinates, the antisymmetry of the
field strength $F_{(p+1)}$ ensures that $\theta_{(p-1)}$ does not
carry a $v$ index. Hence one finds that $i_k\theta_{(p-1)}=0$.

Given the previous results, our final conclusion is that
eq.~(\ref{factor})
applies for an arbitrary product of primary tensors and
Levi-Civita tensors, including arbitrary contractions.

\medskip
\noindent{\it 2) Shift in Tensors}

Next we consider the difference between tensors calculated for the
original and primed metrics.
First the shift in the connection coefficients
${\Omega}^{\mu}_{\nu\lambda}={\Gamma'}^{\mu}_{\nu\lambda}
-{\Gamma}^{\mu}_{\nu\lambda}$ is given by
\beqa\label{omegap}
{\Omega}^{\mu}_{\nu\lambda} &=& \frac{1}{2} \Bigl(
\nabla_{\nu}\bigl(\kappa\, k^{\mu} k_{\lambda}\bigr)
+\nabla_{\lambda}\bigl(\kappa\, k^{\mu} k_{\nu}\bigr)
-\nabla^{\mu}\bigl(\kappa\, k_{\nu} k_{\lambda}\bigr)
\Bigr) \\
&=& {1\over2}\left(k^\mu k_\l\nabla_\nu\kappa+k^\mu k_\nu\nabla_\l\kappa
-k_\nu k_\l\nabla^\mu\kappa\right)
\nonumber\\
&&\quad-{\kappa\over2}\left(k^\mu k_\l\nabla_\nu\AH
+k^\mu k_\nu\nabla_\l\AH-2k_\nu k_\l\nabla^\mu\AH\right)
\eeqa
using eq.~(\ref{w30a}). The corresponding shift in the curvature is then
$R'^{\mu}{}_{\nu\lambda\sigma}
= R^{\mu}{}_{\nu\lambda\sigma} +
\nabla_\lambda \Omega^{\mu}_{\nu\sigma} - \nabla_\sigma
\Omega^{\mu}_{\nu\lambda}
+ \Omega^{\rho}_{\nu\sigma} \Omega^{\mu}_{\rho\lambda}
- \Omega^{\rho}_{\nu\lambda} \Omega^{\mu}_{\rho\sigma}$. However,
the vanishing Lie derivatives, $\cL_k\kappa=0=\cL_k\AH$ 
lead to  $\Omega^{\rho}_{\nu\sigma} 
\Omega^{\mu}_{\rho\lambda}
- \Omega^{\rho}_{\nu\lambda} \Omega^{\mu}_{\rho\sigma} = 0$. 
Hence the shifted curvature reduces to
\begin{equation}\label{riem}
R'^{\mu}{}_{\nu\lambda\sigma} = R^{\mu}{}_{\nu\lambda\sigma} +
\nabla_\lambda \Omega^{\mu}_{\nu\sigma} - \nabla_\sigma
\Omega^{\mu}_{\nu\lambda}
\end{equation}
Hence the curvature with all covariant indices
is $R'_{\mu\nu\lambda\sigma} = R_{\mu\nu\lambda\sigma} +
2 g_{\mu\rho} \nabla_{[\lambda} \Omega^{\rho}_{\sigma]\nu}
+ \kappa\, k_\mu k^\rho R_{\rho\nu\lambda\sigma}$
Now using eq.~(\ref{interest}), as well as rewriting
$2 g_{\mu\rho} \nabla_{[\lambda}\Omega^{\rho}_{\sigma]\nu}$ with
eq.~(\ref{w30a}), we arrive at
\begin{eqnarray}\label{riemans}
R'_{\mu\nu\lambda\sigma} &=& R_{\mu\nu\lambda\sigma}
+ 2 k_{[\mu} \nabla_{\nu]}\nabla_{[\lambda}  \kappa \,k_{\sigma]} 
- 4  \kappa\, k_{[\mu}\nabla_{\nu]}\nabla_{[\lambda}\AH\,k_{\sigma]}
 \nonumber \\
&&+ 3  k_{[\mu} \nabla_{\nu]} \kappa\, k_{[\lambda} \nabla_{\sigma]}\AH
+ 3  k_{[\mu} \nabla_{\nu]}\AH\, k_{[\lambda} \nabla_{\sigma]} \kappa
- 5 \kappa\, k_{[\mu} \nabla_{\nu]}\AH\, k_{[\lambda} \nabla_{\sigma]}\AH
\end{eqnarray}
Here the key observation is that the primed curvature
can be decomposed as $R'_{\mu\nu\lambda\sigma}
=R_{\mu\nu\lambda\sigma}
+ \chi_{\mu\nu\lambda\sigma}$ where the tensor
$\chi_{\mu\nu\lambda\sigma}$ is bilinear in the vector $k_{\mu}$.

In fact, a similar decomposition holds for any covariant derivative
of the curvature. To establish this result, first note that
$\nabla'_\rho R'_{\mu\nu\lambda\sigma} = \nabla_\rho R_{\mu\nu\lambda\sigma}
+ \chi_{\rho\mu\nu\lambda\sigma}$ where
$\chi_{\rho\mu\nu\lambda\sigma}= \nabla_\rho \chi_{\mu\nu\lambda\sigma}
 - \Omega^{\omega}_{\rho\mu}R'_{\omega\nu\lambda\sigma}
 - \Omega^{\omega}_{\rho\nu} R'_{\mu\omega\lambda\sigma}
 - \Omega^{\omega}_{\rho\lambda}R'_{\mu\nu\omega\sigma}
 - \Omega^{\omega}_{\rho\sigma}R'_{\mu\nu\lambda\omega}$.
It is straightforward to show that the tensor
$\chi_{\rho\mu\nu\lambda\sigma}$ is bilinear in the vector $k$.
This is clear for $\nabla\chi$ using eqs.~(\ref{w30a}) and (\ref{riemans}),
and for the $\Omega R$ terms using eq.~(\ref{omegap}). Now similarly 
the $\Omega \chi$ terms are quartic in $k$, but a closer examination
shows that the sum of these terms vanishes. Then, by induction,
we see that if $\chi_{\rho_1 \ldots \rho_n\mu\nu\lambda\sigma}$ is
at least bilinear in $k_\mu$, then
$\chi_{\rho\rho_1 \ldots \rho_n\mu\nu\lambda\sigma}$ must
also be so. This is again straightforward by combining the above
formulae in the definition of $\chi$:
$\chi_{\rho\rho_1 \ldots \rho_n\mu\nu\lambda\sigma}\equiv$
$\nabla'_{\rho} \nabla'_{\rho_1} \ldots \nabla'_{\rho_n}
R'_{\mu\nu\lambda\sigma}
-\nabla_{\rho} \nabla_{\rho_1} \ldots \nabla_{\rho_n} R_{\mu\nu\lambda\sigma}$
$=\nabla_{\rho} \chi_{\rho_1 \ldots \rho_n \mu\nu\lambda\sigma}+$
$\Omega^{\omega}_{\rho\rho_1}\chi_{\omega \ldots \rho_n \mu\nu\lambda\sigma}
+\ldots+$
$\Omega^{\omega}_{\rho\rho_1} \nabla_{\omega} \ldots
\nabla_{\rho_n} R_{\mu\nu\lambda\sigma}+\ldots$.
Hence, all of the new $\kappa$-dependent terms appearing in 
the primed curvature and covariant derivatives thereof,
$\chi_{\rho\rho_1 \ldots \rho_n\mu\nu\lambda\sigma}$,
are at least bilinear in $k$. A detailed inspection shows that
$\chi_{\rho_1 \ldots \rho_n\mu\nu\lambda\sigma}$ contains a sum of
terms of order $k^{2p}$ with $p=1,2,\ldots,1+\lfloor n/2\rfloor$,
where $\lfloor n/2\rfloor$ denotes the integer part of $n/2$.
The precise powers will not be important below,
the key point being that they all begin at $k^2$.

Finally we close this subsection by noting that the Levi-Civita tensor
is invariant under the shift between the original and primed metrics,
\ie $\pol'_{\mu\cdots\rho}=\pol_{\mu\cdots\rho}$. This result
follows since, as noted in section \ref{wavy}, the determinants of the
two metrics are equal because $k$ is null.

\medskip
\noindent{\it 3) Scalar Curvature Invariants}

We are now ready to consider the scalar curvature invariants which 
can be built for the shifted metric. The most general
invariant will consist of an arbitrary product of curvatures, covariant
derivatives thereof\footnote{Here one might also include covariant derivatives
of the scalar $\AH$ which are implicitly geometric tensors
derived from the original metric. Admitting
these extra tensors would not change our conclusions.},
and Levi-Civita tensors, with their indices
(all assumed to be covariant) contracted by the inverse metric
${g'}^{\mu\nu}$.
Hence a generic term is of the form
\begin{equation}\label{invt}
{\cal I'} = \prod^{N}_{j=1} T'_{\nu^j_1\ldots\nu^j_{q_j}} 
\prod^{K}_{k=1}\pol'_{\lambda^k_1\ldots\lambda^k_D}
\prod^{M}_{l=1}g'^{\alpha_l\beta_l}
\end{equation}
along with a rule for contracting the upper
with the lower indices. Here $\sum^{N}_{j=1} q_j + DK = 2M$
where $D$ is the dimension of the spacetime. In fact, one need
only consider $K=0$ or $1$ since the product of two Levi-Civita tensors
can be reduced to a sum of products of metric tensors.

Now from the previous subsection, we know that all of the tensors
in the first product can be decomposed as
$T'_{\mu_1\ldots\mu_n} = T_{\mu_1 \ldots\mu_n} + \chi_{\mu_1 \ldots\mu_n}$
where
$\chi_{\mu_1 \ldots\mu_n}$ is at least bilinear in the vector $k$.
Further, we have $\pol'=\pol$ while the 
inverse metric is given by $g'^{\mu\nu} =g^{\mu\nu} - \kappa k^\mu k^\nu$.
Hence we see that the invariant (\ref{invt}) can be decomposed as follows:
\begin{eqnarray}\label{invt1}
{\cal I'} &=& \prod^{N}_{j=1} \left(T_{\nu^j_1\ldots\nu^j_{q_j}} +
\chi_{\nu^j_1\ldots\nu^j_{q_j}}\right)
\prod^{K}_{k=1}\pol_{\lambda^k_1\ldots\lambda^k_D}
\prod^{M}_{l=1} (g^{\alpha_l\beta_l} - \kappa k^{\alpha_l} k^{\beta_l})  \\
&=& {\cal I} + {\cal J} \nonumber
\end{eqnarray}
where ${\cal I}$ is the invariant of the same algebraic structure as ${\cal
I'}$ but constructed for the original geometry. The difference
${\cal J}$ then contains all of the information about the wave.
Simply multiplying out the terms in (\ref{invt1}),
we may write
\begin{equation}\label{jinv}
{\cal J} = \sum_{i=1} k^{\mu_1} k^{\nu_1}\cdots k^{\mu_i} 
k^{\nu_i}\,{\widehat T}_{\mu_1\nu_1\cdots\mu_i\nu_i}
\end{equation}
where the tensors ${\widehat T}$ are products of primary tensors and 
possibly Levi-Civita tensors, including contractions by $g^{\mu\nu}$.
Given the results of subsection 2), we are assured that ${\cal J}$
is at least bilinear in $k$. Note that antisymmetry of indices,
\eg in $\pol$, may eliminate certain contributions above.
However, from subsection 1), we know that the tensors
${\widehat T}$ factorize as in eq.~(\ref{factor}) when
contracted with $k$. Hence we conclude that 
${\cal J} \propto k_\mu k^\mu = 0$  and so ${\cal I'} = {\cal I}$.
Thus we see that any scalar curvature invariant is identical for
the original and primed metrics, which concludes our proof of the
theorem.

We note that we can easily generalize our theorem to cover scalar
invariants constructed using both the geometry and matter fields.
As in the section \ref{wavy}, we consider a matter sector including
various scalars $\phi_a$, and $p$-form potentials $A_{(p)}$.
We also require that these fields satisfy the same constraints as there:
they are form-invariant along the $k$ flow, \ie
${\cal L}_k \phi_a=0={\cal L}_k F_{(p+1)}$ as in eqs.~(\ref{constr1}) and
(\ref{constr2}). The field strengths
are transverse to the flow, \ie $i_kF_{(p+1)}=k\wedge \theta_{(p-1)}$
as in eq.~(\ref{transverse}). Recall we also have $i_k\theta_{(p-1)}=0$.
Given these results, one can further show that
${\cal L}_k\theta_{(p-1)}=0$ --- this is easily shown by referring to the
local coordinate patches introduced at the end of subsection 1).

Let us then reconsider the contraction identities proved in subsection 1).
Given the properties imposed on the matter fields above, it is
straightforward to extend the discussion to include the scalars,
the field strengths, and covariant derivatives of these, as primary
tensors which satisfy the factorization equation (\ref{factor}).
Similarly one may also show that all higher-order tensors and hence
arbitrary products of the primary tensors satisfy the same factorization
property. Next as in subsection 2), we consider the shift in tensors
calculated for the original and primed metrics. While the scalars
and field strengths themselves are not affected by the shift in the
metric, using eq.~(\ref{omegap}) as well as eq.~(\ref{w30a}), it is
clear that the shifts in covariant derivatives of these fields will
always be at least bilinear in the vector $k$. Hence both of the
crucial results established for the curvature and its covariant derivatives
are easily extended to the matter fields and their covariant derivatives.
Thus the invariant (\ref{invt}) can be extended so that the $T$ also
include these latter fields, and the same final result still holds,
\ie the invariant is independent of the wave profile $\kappa$.

In summary then, we find the rather surprising result that all scalar
invariants, involving any number of covariant derivatives of the curvature
and/or matter fields, are identical for both the original and the
shifted metrics. The essential requirement 
that the original metric has to satisfy is to have a null,
hypersurface-orthogonal, Killing vector, which is supplemented with
certain constraints on the matter fields as well.
Thus no scalar invariant contains any information about the
wave profile $\kappa$, and hence to determine how the geometry has been
modified, one must consider quantities such as tidal forces or 
non-local holonomies. We emphasize that our theorem is of a purely
geometric nature, and holds for
any theory of gravity in any number of dimensions, as long as it
assumes a pseudo-Riemannian geometry as the basis of the description
of gravitational phenomena.

\section{A Five-Dimensional Black String} \label{solex}

Our original motivation in this project was to investigate the properties
of extended black objects in higher dimensions carrying time-dependent or
wavy hair.
The theorem of the previous section tells us that coordinate-invariant
probes are inadequate to examine the properties of such undulating
solutions constructed through the GV technique.
Hence to study the smoothness of the horizon in the presence of
a wave in the next section, we are lead to consider the existence of
parallelly propagated curvature singularities. In order to
do so, however, we must consider a concrete example.
Thus in this section, we present a family of undulating black
strings which are low energy solutions of heterotic string
theory in five dimensions. We begin with a stationary solution with
a null hypersurface-orthogonal Killing vector,
which results from uplifting a four-dimensional
solution first written down by Cveti\v c and Youm \cite{CY}. 
Then we apply the GV technique to generate oscillations on the string.
Similar oscillations of singular strings were considered in
refs.~\cite{DGHW,CMP2}, and of black strings, in refs.~\cite{HM,CT,LW}.

The low energy action for heterotic string theory in five dimensions
includes the following terms
\beq\label{sol16}
I=\int d^5x\sqrt{-G}\e^{-2\Phi}
\left(R(G)+4(\nabla\Phi)^2-{1\over12}{H}^2
-(\nabla\sigma)^2-{1\over4}\e^{2\sigma}{F}^{2}-{1\over4}
\e^{-2\sigma}{\hat{F}}^{2}\right)
\eeq
as well as the metric, we have included
two scalars, the dilaton $\Phi$ and a modulus
field $\sigma$; two gauge fields with $F_{\mu\nu}=
\prt_\mu A_\nu-\prt_\nu A_\mu$ and $\hat{F}_{\mu\nu}=\prt_\mu \hat{A}_\nu
-\prt_\nu \hat{A}_\mu$; and the Kalb-Ramond field with
\begin{equation}
\label{sol17}
H_{\mu\nu\lambda}=\partial_\mu B_{\nu\lambda}
-{1\over2}(A_\mu\hat{F}_{\nu\lambda}+\hat{A}_\mu F_{\nu\lambda})
+({\rm cyclic\ permutations}).
\end{equation}
The metric above is the so-called string-frame metric. The
Einstein-frame metric would be given by
\beq\label{sol18}
g_{\mu\nu}=\e^{-4\Phi/3}G_{\mu\nu}\ .
\eeq
With the latter metric then, 
the dilaton coupling in the Einstein term is eliminated and
the action becomes
\beqa\label{sol19}
I&=&\int d^5x\sqrt{-g}
\left(R(g)-{4\over3}(\nabla\Phi)^2-{1\over12}\e^{-8\Phi/3}{H}^2
\right.\nonumber\\
&&\qquad\qquad\qquad\qquad\qquad\left.
-(\nabla\sigma)^2-{1\over4}\e^{2\sigma-4\Phi/3}{F}^{2}-{1\over4}
\e^{-2\sigma-4\Phi/3}{\hat{F}}^{2}\right)\ .
\eeqa
However, we choose to present our solution in terms of the
string-frame metric, which has a much simpler appearance in the
present case.

We will be interested in the following black string solution:
The string frame metric is 
\begin{equation}
\label{sol11}
ds^2={f\over h}\,du^2
+{2\over h}\,du\,dv+{k\ell}\,\left(dr^2+r^2(d\theta^2
+\sin^2\theta\,d\phi^2)\right)
\end{equation}
while the remaining fields are given by:
\beqa\label{sol12}
B&=&{1\over h}\,du\wedge dv
\nonumber\\
A&=&-P_1\cos\theta\,d\phi \qquad
\hat{A}=P_2\cos\theta\,d\phi
\nonumber\\
\e^{2\sigma}&=&\ell/k\qquad\qquad\qquad
\e^{4\Phi}=k \ell/h^2\ .
\eeqa
where we defined the following functions:
\begin{eqnarray}\label{sol10}
f&\equiv&1+{Q_1\over r}\qquad\qquad
h\equiv1+{Q_2\over r}\nonumber \\
k&\equiv&1+{P_1\over r}
\qquad\qquad \ell\equiv 1+{P_2\over r}\ \ .
\end{eqnarray}
Hence this configuration is specified by
four different parameters. The solution could be simplified
by setting all of these equal, however, we wish to illustrate
that our results apply in the generic case. We will assume that
all of the constants are positive, in order that our solution
properly describe a black string with a horizon at $r=0$.

Considering the asymptotic metric, one has for large $r$
\begin{equation}
\label{sol20}
ds^2\rightarrow du^2  +{2}\,du\,dv+dr^2+r^2d\Omega =-dt^2+dy^2+
dr^2+r^2d\Omega
\end{equation}
where $y=u+v$ and $t=v$. Hence we should consider $y$ as the spatial
coordinate running parallel to the string, while
$t$ is the asymptotic time. 
Note that as $r\rightarrow0$, $g_{rr}\simeq{P_1P_2\over r^2}$
indicating the presence of a degenerate horizon.
Near the horizon, the metric becomes
\begin{equation}
\label{sol22}
ds^2\simeq {Q_1\over Q_2}du^2  +{2r\over Q_2}\,du\,dv+P_1P_2
\left[\left({dr\over r}\right)^2+d\Omega^2\right]
\end{equation}

The solution has two Killing vectors which are of interest
\begin{equation}
\label{sol23}
k^\mu\,\partial_\mu=\partial_v=\partial_t+\partial_y
\qquad\qquad h^\mu\,\partial_\mu
=\partial_u=\partial_y\ \ .
\end{equation}
(These are in addition to the standard rotational Killing vectors
for $\theta$ and $\phi$.) The first of these is the null generator
of the horizon. In this role, $k$ has the rather unusual feature 
that it is null everywhere -- not just at the horizon.
Further, it is not given by
$\partial_t$ in the asymptotic coordinates, rather we have
$k^\mu\prt_\mu=\partial_v=\partial_t+\partial_y$.
Here the $\partial_y$ contribution is related to the presence
of linear motion along the $y$ direction.
The coefficient may be interpreted as the `horizon velocity', which in the
present case is one, \ie the speed of light.

The null Killing vector $k$ is hypersurface-orthogonal as well,
satisfying
\begin{equation}
\label{sol24}
\nabla_{[\mu} k_{\nu]}=k_{[\mu}\nabla_{\nu]}\log h
\end{equation}
where $h$ is the same function defined in eq.~(\ref{sol10}).
Hence the metric admits the symmetry desired for the wave-generating
technique. It is also straightforward to show that the matter fields
satisfy the appropriate conditions (\ref{constr1}--\ref{transverse}).
Hence following the discussion of section \ref{wavy}, we apply 
eq.~(\ref{w311}) to define a new string-frame metric
\begin{equation}
\label{sol25}
G'_{\mu\nu}=G_{\mu\nu}+h\,\Psi\,k_\mu k_\nu
\end{equation}
where following eq.~(\ref{psicon2}), $\Psi$ is chosen to satisfy
\begin{equation}
\label{sol26}
k^\mu\nabla_\mu\Psi=0,
\qquad{\rm and}\qquad
\partial_\mu\left(\e^{-2\Phi}\sqrt{-G}G^{\mu\nu}\partial_\nu\Psi\right)=0\ .
\end{equation}
The first condition is simply $\partial_v\Psi=0$, {\it i.e.,}
$\Psi$ is independent of $v$. In the present case, 
the second condition reduces to
$\nabla_F^2\Psi=0$, \ie Laplace's equation on a flat spatial metric in
the transverse coordinates ($r,\theta,\phi$). Thus the general solution
for eq.~(\ref{sol26}) may be written as
\beq\label{Psol}
\Psi=\sum_{l,m}\left(a_{lm}(u)\,r^l+b_{lm}(u)\,r^{-(l+1)}\right)
Y_{lm}(\theta,\phi)
\eeq
where $Y_{lm}(\theta,\phi)$ are usual spherical
harmonics, and $a_{lm}$ and $b_{lm}$ are arbitrary functions of $u$.

Let us consider the various perturbations in turn.
The case with $r^l$ and $l=0$ yields
an asymptotically flat metric, which in fact is
diffeomorphic to the original,
\begin{eqnarray}
\label{sol28}
d{s'}^2&=&
{1\over h}\left(
f+a(u)\right)\,du^2
+{2\over h}\,du\,dv+{k\ell}(dr^2+r^2d\Omega) \nonumber \\
&=&{f\over h}\,du^2
+{2\over h}\,du\,d\tilde{v}+{k\ell}(dr^2+r^2d\Omega)
\end{eqnarray}
where $\tilde{v}=v+{1\over2}a(u)du$. Note that the
constant term in $f$ is a special constant case of these perturbations,
and so could also be eliminated in the same way.
With the choice $r^l$ and $l>1$,
the metric is not asymptotically flat, and so we do not consider these
solutions as providing perturbations intrinsic to the black string,
\ie `wavy' hair. Instead
they would be more accurately described as embedding the string
in a space filled with (asymptotic) gravitational radiation.
The same is apparently true for the $r^l$ mode with $l=1$, but in fact this
solution yields an asymptotically flat metric, as is seen as
follows \cite{DGHW,CMP2}:
First introduce Cartesian coordinates on the transverse space, in
which case the wavy metric becomes
\beq\label{sol82}
d{s'}^2={1\over h}\left(f+a_i(u)x^i\right)\,du^2
+{2\over h}\,du\,d{v}+{k\ell}\,dx_idx^i\ .
\eeq
However the following coordinate transformation,
\beqa\label{cotrans}
\tilde{v}&=&v+\dot A_ix^i-{1\over2}\int^u\dot A_i^2du \nonumber \\
\tilde{x}^i&=&x^i-A^i
\eeqa
with $2\ddot A_i\equiv a_i$, produces a metric which is manifestly
asymptotically flat
\beq\label{flat}
d\tilde{s}^2={1\over h}\left(
f+(hk\ell-1)\dot A_i^2\right)\,du^2
+{2\over h}\,du\left(d\tilde{v}+(hk\ell-1)\dot A_id\tilde{x}^i\right)
+{k\ell}\,dx_idx^i\ .
\eeq
These waves represent oscillations of the string in the transverse space.

The perturbations generated with $r^{-(l+1)}$
are all localized near the horizon,
and leave the metric asymptotically flat. Hence we may consider
these deformations as candidates for `wavy' hair on the black string.
For $l>0$, these deformations produce $g_{uu}\rightarrow\infty$
as we approach the `horizon' at $r=0$.
Note however that this divergence does not effect the volume element,
$\sqrt{-g}$, and so we should be careful in deciding
whether or not these perturbations produce a true singularity at 
$r=0$.\footnote{The divergence in $g_{uu}$ does indicate a
divergence of the norm of the Killing vector $h^\mu\prt_\mu=\prt_u$,
which could be interpreted as a geometric singularity \cite{HM2}.
One should still demonstrate that $r=0$ is accessible to causal observers,
\ie that this region `belongs' to the spacetime, as is done in our
following analysis.}
In the next section, we will see that in fact
with $l>0$ these solutions are singular, and hence
we are only left with a wavy black string for $l=0$ in which case
\beq\label{sol29}
d{s'}^2={1\over h}\left(f+{b(u)\over r}\right)\,du^2
+{2\over h}\,du\,dv+{k\ell}(dr^2+r^2d\Omega) 
\eeq
These perturbations represent longitudinal waves carrying momentum
along the string without oscillations. Note that the $Q_1$ term in $f$
represents a constant contribution to $b(u)$. 
One may wonder if this wave is really physical or merely an artifact 
of an awkward choice of coordinates, given our theorem on the elusiveness
of the wave profile. To see that it is
indeed physical, we can compute its mass per unit length.  Since the
oscillating string (\ref{sol29}) is asymptotically flat, and we can use the
coordinates of eq. (\ref{sol20}), we can determine the mass per unit length
according to $dE/dy = -(1/8\pi) \int_S \pol_{ybcde} 
dx^b dx^c \nabla^d \xi^e$, where all the tensors are defined in the Einstein 
frame. The vector field $\xi=\partial_t$ here is the asymptotic generator of
time translations, and the integration is carried over a sphere
at spatial infinity. The result is 
$dE/dy = (P_1 + P_2 + Q_2 + 3Q_1 +3b(y-t))/6$ and since the mass per unit
length depends on the wave profile, we see that the solution is really
a superposition of the string and the wave and so is clearly different form 
the stationary string, where $b=0$. 

\subsection{Parallelly Propagated Singularities} 

Having applied the GV technique to the original solution above,
we have apparently generated a large family of oscillating
black string solutions. However, 
as discussed above, we have reason
to worry that some of the modes may actually introduce a
curvature singularity at the null surface which was originally the
black string's horizon. Normally the approach to proving the existence of
a horizon would be to find coordinates in which the metric is analytic
at the null surface in question. For the present undulating solutions,
finding such coordinates is an enormous problem (see, \eg \cite{HM}).
While we address this question for the monopole waves, \ie those
with $l=m=0$, in the following section, the task at hand is in fact
much simpler. We wish to show that for $l>0$ a given null hypersurface is not
a horizon, which simply requires finding any geometric quantity
which diverges when the surface is approached along some geodesic.
By the properties of our original black string and the approach
used to generate their wavy counterparts,
the theorem of section \ref{long} tells us that no scalar invariants
involving the curvature and/or the matter fields
contains any information about the waves.
Therefore given that all scalar invariants are insensitive to the oscillations
and thus to any singularity which they may introduce,
we resort to alternative means of probing the wavy geometry.

Tidal forces prove to be a good tool for resolving our problem.
If we approach the null surface along a geodesic, and 
we allow our observer to be slightly non-local (\eg a
string which may indeed seem the natural probe in the present context),
this observer will be able to
determine the differences between the gravitational forces acting at different 
points. These forces are determined by the Riemann curvature measured in the
rest frame of our observer. One may object to the notion of the observer's
rest frame, as we have just said that the observer of interest is
non-local. We will
assume that the extension of the observer is controllably small, and hence that
the center-of-mass frame represents a good reference frame in which
to express the results.

Still, identifying a convenient geodesic trajectory to follow
in the presence of a general oscillation proves to be beyond our abilities.
The reason is that one cannot find enough
integrals of motion to solve the problem in terms of quadratures. We
therefore restrict our attention to wave profiles which are constant in $u$.
In this case the extra Killing vector, $h^\mu\prt_\mu=\prt_u$,
yields an additional constant of the motion, enabling
us to find analytically suitable 
geodesics for any mode. These solutions should be a very good
approximation for backgrounds with slowly varying wave profiles.
When we compute
the full curvature of an undulating string, we can see that any $u$
dependence only adds contributions of a subleading order. Hence 
if the curvature turns out to be divergent as some hypersurface $r = constant$
is approached, the constant
profile solutions will contain all the information about the leading order of
divergences.

We present our calculations in several steps. First we will  
show that for each mode of oscillation, there exists a geodesic stretching 
between the null surface and the asymptotic infinity --- hence showing
that both of these regions belong to the space time.
Next we will construct the Lorentz transformation relating a natural
stationary orthonormal frame to the
rest frame of the observer moving along the geodesic. Finally we will
consider the orthonormal frame curvature and boost it to the
frame of the infalling observer, in order to find the tidal forces
which he measures. The divergences found in this way are equivalent to
parallelly propagated curvature singularities.
We will isolate the leading divergences of the tidal forces for all modes
both at $r=0$ and asymptotically, finding that all the
localized multipoles with $l\ge1$ 
have unbounded tides on the null surface and that all
the growing multipole modes with $l\ge2$ have divergences at asymptotic
infinity.

\medskip
\noindent{\it 1) Geodesics}

We begin by examining the timelike geodesics of a wavy solution which
is excited by a single mode.
We will demonstrate that there are always
geodesics extending between $r=0$ and the asymptotic region
$r\rightarrow\infty$, and the former is reached in finite
affine time when starting from finite $r$.
Hence this null surface $r=0$ must be included in the manifold
described by the wavy solution.

We may write the general solution as 
\beq\label{metro}
ds^2 = 2 F_2 du dv + F_3^2 du^2 
+ \frac{1}{F_1^2}\left(dr^2+r^2(d\theta^2+\sin^2\theta\,d\phi^2)\right)
\eeq
where
\beqa
F_2 &=& \frac{1}{h} \qquad\quad F_1^2 = \frac{1}{kl}
\qquad\quad F_3^2 = \frac{f+\Psi}{h}\nonumber\\
{\rm and}\quad \qquad\Psi &=& B(u) r^{\beta}
P^m_l(\cos\theta) \cos \left(m \phi+\delta(u)\right)
\label{constprof}
\end{eqnarray}
and $f,h,k,l$ are defined in eq.~(\ref{sol10}).
To produce a simple real metric, we have expressed the
angular dependence in terms of an associated Legendre
function of the first kind, $P^m_l (\cos\theta)$, as well as
the $cos(m\phi+\delta)$ factor, rather using spherical harmonics
as in eq.~(\ref{Psol}).
Those solutions (\ref{Psol}) are then reproduced by setting
$\beta= -(l+1)$ or $l$, which corresponds to what we will call 
the localized and the growing modes, respectively. However, for much
of the following analysis, we will leave this exponent as $\beta$
in order to emphasize the contributions coming from the differentiation
of this factor. In principle the amplitude $B$ and the phase
$\delta$ are arbitrary functions of $u$, but as discussed above,
to simplify the analysis of the geodesics, we will set both of
these to be constants in the following. This will be enough to 
identify the leading divergences, and should still provide good
approximation in the case of a slow $u$ dependence. In fact we will set
$\delta=0$, which can be attained with a simple shift of $\phi$.

As usual to obtain the geodesic equations, we simply consider the Lagrangian
$L = (ds/d\lambda)^2$, and write down the Euler-Lagrange equations. Because
the Lagrangian does not
contain a potential, the effective Hamiltonian is conserved, giving
$(ds/d\lambda)^2 = const$. Given
this integral of motion, we need not consider 
the Euler-Lagrange equation for the radial coordinate. In addition, the
two translational Killing vectors, $\partial_u, \partial_v$,
produce two more first integrals, which leaves us with second
order differential equations only for the two angular coordinates.
Hence the equations for timelike geodesics are
\begin{eqnarray}
\label{geod}
&&F_2 u' = p-\omega ~~~~~~~~~  F_2 v' + F_3^2 u' = p
~~~~~~~~~ \nonumber \\
&&2 F_2 v' u' + F_3^2 u'^2 + \frac{1}{F_1^2} 
\bigl(r'^2 + r^2 \theta'^2 + r^2
\sin^2\theta \phi'^2 \bigr) = -1 \nonumber\\
&&2\bigl(\frac{r^2 \theta'}{F_1^2}\bigr)' = 
2 \frac{r^2 \sin\theta \cos\theta}{F_1^2} \phi'^2
+ \frac{B}{h} r^{\beta} 
\frac{d P^m_l(\cos \theta)}{d\theta} \cos m \phi ~u'^2 \\
&&2(\frac{r^2\sin^2\theta}{F_1^2} \phi')' = - m \frac{B}{h} r^\beta
P^m_l (\cos\theta) \sin m \phi ~u'^2  \nonumber
\end{eqnarray}
where we expressed the integrals of motion as $\omega$ and $p$.
When $\beta<0$, one finds that in the asymptotic region (\ref{sol20})
these correspond to the energy (which is assumed to be positive),
and linear momentum along $y$, respectively. The last of
these equations may be solved by setting $\phi=\phi_n={\pi\over m}n$ with
$n=0,1,\ldots,2m-1$ for $m\ne0$, while for $m=0$
we may fix $\phi$ to be any constant.
Now with the constant $\phi$, the first term on the RHS of
the second-to-last equation also vanishes.
Hence the latter equation is solved if we now choose $\theta=\theta_0 =
constant$ where $\theta_0$ corresponds to an extremum of
Legendre function $P^m_l(\cos \theta)$.
Then the independent set of equations
defining these radial geodesics may be written,
using $F_3^2 = ({f}+(-)^n{B} r^{\beta} P^m_l(\cos\theta_0))/h$,
\begin{eqnarray}
&&u' =  \frac{p-\omega}{F_2}  ~~~~~~~~~   v'  = \frac{p}{F_2} + \left(
{f}+(-)^n{B} r^{\beta} P^m_l(\cos \theta_0)\right)
\frac{\omega-p}{h\,F_2^2} \label{geod1} \\
&& r'^2 = \left({f}+(-)^n  {B} r^{\beta} 
P^m_l(\cos \theta_0)\right)  
\frac{F_1^2}{h\,F_2^2} (\omega-p)^2 + 2 \frac{F_1^2}{F_2} (\omega-p) p - F_1^2
\label{geod2}
\end{eqnarray}
Now we must
choose $\theta_0$ and $\phi_n$, as well as the constants
$\omega$ and $p$, in such a way that our geodesics
(a) extend to infinity and (b) reach the null
surface $r=0$. We need the former to insure
that our observer is physically connected
to recording devices infinitely far away from the gravitational source.
The second condition that the geodesic does not turn before reaching $r=0$
is necessary because we want to probe this region for singularities.

We begin by considering the localized modes with $\beta=-(l+1)<0$.
From the radial equation (\ref{geod2}), we see that as $r \rightarrow \infty$,
for which $h,\,F_{1,2} \rightarrow1$ and $r^\beta\rightarrow0$, so
$r'^2 \rightarrow \omega^2-p^2 -1$. Thus we require $\omega\ge\sqrt{p^2+1}$
so that the geodesics extend to infinity. Now
 in the limit $r \rightarrow 0$, we have $F_2=h^{-1}\propto r$,
$F_1^2\propto r^2$ and $f\propto 1/r$. Thus the leading contribution
on the RHS of eq.~(\ref{geod2}) comes from the $r^\beta$ term when
$\beta<-1$, and hence we must choose $\theta_0$ and $n$ such that
$(-)^nB P^m_l(\cos \theta_0) >0$ in order that no turning points
occur before reaching $r=0$. For $l\ge1$ there exist many extrema
of $P^m_l(\cos \theta)$, and it is straightforward to verify that
one has enough freedom to choose the angles in order that the
geodesic reaches $r=0$ (even for the case $m=0$). For the special
case $\beta=-1$ or $l=0$, turning points are evaded as long as
$Q_1+B>0$.\footnote{Note then that for $l=0$ and $B<-Q_1$, one has
solutions for which the radial geodesics can never reach $r=0$.
As pointed out in ref.~\cite{HGT}
however, when the $y$ direction is compactified in this case, there appear
closed timelike curves, which can communicate with observers at
infinity, implying the breakdown of chronology.
In any event, we will ignore this pathological case in the following.}

For the growing modes with $\beta = l\ge0$, the analysis is similar to that
above. Note that in this case, the interpretation of $\omega$ and $p$
would not be correct since the asymptotic structure
of the metric is modified. Beginning with the limit $r\rightarrow 0$,
we can ignore $r^\beta$ contribution in eq.~(\ref{geod2}) and in
fact one finds that this region is always reached without imposing
any constraints. 
In the asymptotic region $r\rightarrow \infty$, the $r^\beta$ term
dominates eq.~(\ref{geod2}) for $\beta>0$, and we must again
choose the angles such that
$(-)^nBP^m_l(\cos\theta_0) >0$ if the geodesics are to extend all
the way to infinity without turning back. For the special case
$\beta=0=l$, the constraint to reach the asymptotic region becomes
$(\omega-p)(\omega+p+B)\ge1$ which can always be satisfied with 
an appropriate choice for the integrals of the motion.

In the following, it will be convenient to absorb $(-)^nP^m_l(\cos \theta_0)$
into the amplitude. Hence we define $\B\equiv (-)^nBP^m_l(\cos \theta_0)$.
For the special case with $\beta=-1$ and $l=0$, we set
$\B\equiv Q_1+B$.

\medskip
\noindent{\it 2) Lorentz Transformation}

Hence we have defined an interesting set of radial geodesics
for which the tangent vector is given by $V^\mu=dx^\mu/d\lambda$ where
\beqa\label{compost}
&&u' = {p-\omega\over F_2}
\qquad\quad
v'  = {p\over F_2} + (f +  \B r^{\beta}){\omega-p\over h\, F_2^2}
\nonumber\\
&&r' = \pm \left((f +  \B r^{\beta}) {F^2_1\over h\, F^2_2}(\omega-p)^2
+ 2 {F^2_1\over F_2} (\omega-p)p - F_1^2 \right)^{1\over2}
\eeqa
and $\theta'=0=\phi'$. Here the $-$ ($+$) sign corresponds to inward
(outward) directed geodesics. We will want to examine the tidal forces
in the rest frame of an observer moving with this five-velocity.
Hence as an intermediate step, we determine 
the Lorentz transformation which takes us from a natural stationary
frame, in which the curvature is easily calculated, to the
observer's freely-falling frame. 

First to define our stationary orthonormal frame, we
complete the squares in our general metric (\ref{metro})
\beq\label{metro2}
ds^2 =-{F^2_2\over F^2_3}dv^2+ F_3^2 (du^2 + \frac{F_2}{F_3^2} dv)^2
+ \frac{1}{F_1^2}\left(dr^2+r^2(d\theta^2+\sin^2\theta\,d\phi^2)\right)
\eeq
We see an obvious orthonormal basis of one-forms is
\begin{equation}
\label{funf}
e^0 = \frac{F_2}{F_3} dv
\qquad 
e^{4} = F_3 du + \frac{F_2}{F_3} dv
\qquad
e^r = \frac{dr}{F_1}
\qquad
e^\theta = \frac{r\,d\theta}{F_1}
\qquad
e^\phi = \frac{r\sin\theta\,d\phi}{F_1}
\end{equation}
Note that in this basis, $e^0$ is distinguished as the unit time-like
one-form, at least everywhere along our radial geodesics.\footnote{This
is guaranteed since $F^2_2=h^{-2}$ is trivially positive, while
we ensured that $F_3^2>0$ in order that 
the geodesics reached between $r=0$ and $r\rightarrow\infty$.
The sole exception, which we ignore, is the case $\beta=0=l$
for which we could have $F_3^2<0$ if $B<-1$.}
In this frame, we have  $V^a = e^a{}_\mu V^\mu$ where the
$e^{a}{}_\mu$ are the components of the f\"unfbein (\ref{funf})
\begin{equation}
\label{4vel}
V^0 =  \frac{ F_3}{F_2}(\omega-p) +\frac{p}{F_3}
~~~~~~~~~~~  V^4 = \frac{p}{F_3}
~~~~~~~~~~~  V^r = \pm \left(\frac{F_3^2}{F_2^2}(\omega-p)^2 
+ 2\frac{(\omega-p) p}{F_2} -1\right)^{1\over2}
\end{equation}
along with $V^\theta=0=V^\phi$.
As a check, one may easily verify that $\eta_{ab}V^a V^b=-1$.

Now we wish to find a Lorentz transformation which takes a unit time-like
vector $N^a=\delta^a_0$ into the observer's five-velocity:
$V^a=L^a{}_bN^b$. Then applying this transformation to our stationary
f\"unfbein (\ref{funf}) would produce a natural basis of orthonormal
one-forms which the observer might use in his rest frame. Of course,
we are left with some ambiguity in the choice of the SO(1,4) matrix
defining this Lorentz transformation. One approach to resolving this
ambiguity is defining the transformation by parallelly propagating
the stationary frame in from infinity along our radial geodesic.
Since $V^\theta=0=V^\phi$, a much less labor-intensive approach is to
simply take the uniquely defined SO(1,2) matrix mixing only the 0, 4
and $r$ directions. The latter will differ from that defined through
parallel propagation by an SO(4) transformation mixing the spatial
one-forms. Such a rotation however will not introduce any new divergences,
and hence it suffices to consider the simpler boost. One can think
of this transformation as the result of parallelly propagating in
not our stationary f\"unfbein (\ref{funf}) but rather a rotated version
of it. The simpler SO(1,2) Lorentz transformation may be written as
\begin{equation}
\label{loren}
L^a{}_b=~\pmatrix{
V^0 & V^4& V^r & 0 &0\cr
V^4 & 1+{(V^4)^2\over V^0(V^0+1)} & {V^4\,V^r\over V^0(V^0+1)}& 0 &0\cr
V^r & {V^4\,V^r\over V^0(V^0+1)}& 1+{(V^r)^2\over V^0(V^0+1)} &0 &0\cr
0 & 0 & 0 & 1 & 0 \cr
0 & 0 & 0 & 0 & 1 \cr}
\eeq

At this point let us comment at the behavior of the five-velocity
in the regions of interest. For $\beta\le-1$ and $r\rightarrow0$,
the dominant factor is $F_3\propto r^{(\beta+1)/2}$ leading to
$V^0\simeq (\omega-p)\sqrt{\B Q_2}\, r^{(\beta-1)/2}\simeq V^r$ 
while $V^4\simeq0$. Hence the observer is accelerated 
to almost a null radial geodesic as he
nears $r=0$, and $L$ approaches an infinite boost in the radial direction.
For $\beta\ge0$, $F_3$ and all of the components of $V^a$
are finite at this null surface.
Similarly for $\beta>0$ and $r\rightarrow\infty$, $F_3\propto r^{\beta/2}$
again dominates to producing the five-velocity $V^0\simeq
(\omega-p)\sqrt{\B} r^{\beta/2}\simeq V^r$ and $V^4\simeq0$.
Hence eq.~(\ref{loren})
yields an infinite radial boost as $r\rightarrow\infty$. Finally, we
note that for $\beta\le1$ and $\beta=0$, $F_3$ and all of the
components of $V^a$ and $L$
remain finite in the asymptotic region.

\medskip
\noindent{\it 3) Divergent Tides}

Now we wish to examine the tidal forces experienced by an observer
following our radial geodesics. 
We will focus here on the curvature components with
all indices in the $0,4,r$ subspace, and study the corresponding
components in the observer's rest frame. Since our Lorentz
transformation (\ref{loren}) does not mix these indices with the
$\theta, \phi$ directions, it simplifies our presentation to
only consider these components.
Our omission of the components carrying angle indices does not
imply that they are finite. In fact the components with two indices
in the $\theta,\phi$ subspace also typically diverge whenever
we find divergences in the following analysis, and in
qualitatively the same way, as discussed at the end of this section.
Our final results are that divergent gravitational tides appear
on the null surface $r=0$ for the localized modes with $\beta<-1$
and at asymptotic infinity for the growing modes with $\beta>2$.

With our definition of the orthonormal basis (\ref{funf}),
it is straightforward to compute the curvature in this frame.
As discussed above, we focus on the curvatures in the $0,4,r$ subspace
for which the nonzero components are
\begin{eqnarray}
\label{curv}
R^{4r4r}&=&
- \frac{F_1}{F_3} F'_3\,F'_1
+\frac{F^2_1}{F_2 F_3} F'_3\,F'_2 
-\frac{F_1^2}{F_3} F''_3
-\frac{F_1^2}{F_3^2} (F'_3)^2
-\frac{F^2_1}{4F^2_2} (F'_2)^2 \nonumber \\
R^{0r0r} &=&
- \frac{F_1}{F_3} F'_3\,F'_1
+\frac{F^2_1}{F_2F_3} F'_3\,F'_2
-\frac{F_1^2}{F_3} F''_3 
- \frac{F_1^2}{F_3^2} (F'_3)^2 \nonumber \\
&&\qquad \qquad \qquad \qquad +\frac{F^2_1}{F_2} F''_2 
+\frac{F_1}{F_2}F'_2\,F'_1
- \frac{3F_1^2}{4F_2^2} (F'_2)^2 \\
R^{0404} &=& \frac{F_1^2}{4F_2^2} (F'_2)^2 
\nonumber\\
R^{4r0r}&=& 
-\frac{F_1}{F_3} F'_3\,F'_1
+\frac{F^2_1}{F_2F_3} F'_3\,F'_2
- \frac{F_1^2}{F_3} F''_3
-\frac{F_1^2}{F_3^2} (F'_3)^2\nonumber\\
&&\qquad \qquad \qquad \qquad +\frac{F^2_1}{2F_2} F''_2 
+\frac{F_1}{2F_2} F'_2 F'_1
- \frac{F^2_1}{2F_2^2} (F'_2)^2
\nonumber
\end{eqnarray}
where the primes indicate partial derivatives with respect to $r$.

Now we wish to consider $r\rightarrow0$. To simplify our calculations
further, we only consider the leading order of these curvature components.
This approach is consistent, because we will see that all of the nonzero
components (\ref{curv}) are of the same order of
magnitude at the null surface. In fact these terms are all finite there,
and so any divergence in the tidal forces can only arise from the
boost matrix (\ref{loren}) upon transforming to the observer's rest
frame. Given the discussion at the end of the previous subsection then,
it is clear that the only possible divergences will occur for
the modes $\beta=-(l+1)<0$. For these modes, as $r\rightarrow0$,
$F_1 \rightarrow r/\sqrt{P_1 P_2}$, 
$F_1' \rightarrow 1/\sqrt{P_1 P_2}$, $F_1'' \rightarrow 
-(P_1 + P_2)/(P_1 P_2)^{-3/2}$,
$F_2 \rightarrow r/Q_2$, $F_2' \rightarrow 1/Q_2$, $F_2'' \rightarrow -2/Q_2^2$
and
$F_3 \rightarrow \sqrt{B_0/Q_2} r^{-l/2}$,
$F_3' \rightarrow - (l/2)\sqrt{B_0/Q_2} r^{-l/2-1}$ and 
$F_3'' \rightarrow (l(l+2)/4)\sqrt{B_0/Q_2} r^{-l/2-2}$,
and so the above curvature components (\ref{curv}) reduce to
\begin{eqnarray}
\label{curvapp}
R^{4r4r} &\rightarrow& - \frac{1 + 2 \beta(\beta+1)}{4 P_1 P_2} ~~~~~~~~~~
R^{0r0r}\rightarrow \frac{1 - 2 \beta(\beta+1)}{4 P_1 P_2} \nonumber \\
R^{0404} &\rightarrow& \frac{1}{4 P_1 P_2} ~~~~~~~~~~ R^{4r0r} \rightarrow
-\frac{\beta(\beta+1)}{2 P_1 P_2}
\end{eqnarray}

Transforming to the observer's rest frame, 
$\hat R^{abcd} = L^{a}{}_k L^{b}{}_l L^{c}{}_m L^{d}{}_n R^{klmn}$, we find
\begin{equation}
\label{curvappboost}
\hat R^{abcd} \rightarrow 
\frac{1}{4 P_1 P_2} \{ \Delta^{abcd}_1 - 2 \beta(\beta+1)
\Delta^{abcd}_2 \}
\end{equation}
where
\begin{eqnarray}
\label{deltas}
\Delta_1^{4r4r} &=& -1 \qquad \qquad \qquad\qquad \qquad \Delta_1^{0r0r} = 1
\nonumber\\
\Delta_1^{0404} &\rightarrow& 1\hphantom{-} \qquad \qquad \qquad\qquad \qquad 
\Delta_1^{4r0r} = 0   \\
\Delta_2^{4r4r} & \rightarrow& (V^0)^2\simeq (\omega-p)^2\B Q_2\,r^{\beta-1}
\qquad\quad
\Delta_2^{0r0r} \rightarrow 1 \nonumber \\
\Delta_2^{0404} &\rightarrow&(V^r)^2\simeq (\omega-p)^2\B Q_2\,r^{\beta-1}
\qquad\quad
\Delta_2^{4r0r}\rightarrow V^4(V^r)^2\simeq p(\omega-p)^2\sqrt{\B Q_2^3}\,
r^{(\beta-3)/2}\nonumber
\end{eqnarray}
Hence we see that there are divergences but that they only appear in the
terms proportional to $\beta(\beta+1)$. The $\beta$-independent 
contributions are essentially boost invariant. These divergences are
therefore absent for $\beta=-1$ or $l=0$. This result might have been
expected since in this case, for a constant
wave profile, the solution is essentially the original regular black string
with a modified $Q_1$.
The leading divergences are easily seen to be
\begin{eqnarray}
\label{divs}
\hat R^{4r4r} &\rightarrow& - \frac{l (l +1)}{2} 
\frac{\B Q_2 (\omega-p)^2}{P_1 P_2 r^{l+2}}
~~~~~~~~~~
\hat R^{0r0r} \rightarrow \frac{1 - 2 l(l+1)}{4 P_1 P_2} \\
\hat R^{0404} &\rightarrow& - \frac{l (l +1)}{2}
\frac{\B Q_2(\omega-p)^2}{P_1 P_2 r^{l+2}}
~~~~~~~~~~
\hat R^{4r0r} \rightarrow - \frac{l (l +1)}{2}
\frac{\sqrt{\B Q_2^3} (\omega-p)^2 p}{P_1 P_2 r^{l/2 +2}}
\nonumber
\end{eqnarray}
So we see that for all higher multipoles with $\beta=-(l+1)<-1$ or $l>0$,
there appear singular tidal forces on the null surface $r=0$.
Because these divergences will not be cancelled by any
other terms of the metric for slowly oscillating strings, we 
conclude that all these space-times have a null singularity at $r=0$.
Hence, the excitation of these higher modes on the black string results
in the appearance of naked singularities,
and so the resulting solutions are no longer black strings after all.
Thus we rule out all of these higher multipoles as a variety
of non-stationary hair.

A similar calculation shows that the growing wave modes with $\beta=l>1$ have
diverging tidal forces in the limit $r\rightarrow \infty$. In this limit,
$F_1 \rightarrow 1$, 
$F_1', F_1'' \rightarrow 0$,
$F_2 \rightarrow 1$, $F_2', F_2'' \rightarrow 0$
and
$F_3 \rightarrow \sqrt{B_0} r^{\beta/2}$,
$F_3' \rightarrow (\beta/2)\sqrt{B_0} r^{\beta/2-1}$ and 
$F_3'' \rightarrow (\beta(\beta-2)/4)\sqrt{B_0} r^{\beta/2-2}$, 
and hence to the leading order, the curvature components (\ref{curv})
become
\begin{equation}
\label{curvappinft}
R^{4r4r} =  R^{0r0r} = - R^{4r0r} \rightarrow - \frac{\beta(\beta-1)}{2 r^2}  
\end{equation}
The boosted curvatures, to the leading order, are given by
\begin{equation}
\label{curvappboost1}
\hat R^{abcd} \rightarrow 
-\frac{\beta(\beta-1)}{2r^2}\Delta^{abcd}_2
\end{equation} 
The terms proportional to $\Delta^{abcd}_1$ are all of the subleading order.
The limiting values of 
$\Delta^{abcd}_2$ when $r\rightarrow \infty$ are given by
\beqa
\label{deltasinft}
\Delta_2^{4r4r} \rightarrow (\omega-p)^2 B_0 r^\beta
&\qquad& \Delta_2^{0r0r} \rightarrow 1 \\
\Delta_2^{0404} \rightarrow (\omega-p)^2 B_0 r^\beta &\qquad&
\Delta_2^{4r0r} \rightarrow (\omega-p + p (\omega-p)^2) B_0 r^{\beta/2}
\nonumber
\eeqa
and so the leading divergences are easily seen to be
\beqa
\label{curvappinftboost}
&& \hat R^{4r4r} = \hat R^{0r0r} \rightarrow 
- \frac{\beta(\beta-1)}{2}(\omega -p)^2 B_0 r^{\beta-2} \nonumber \\
&& \hat R^{4r0r} \rightarrow 
\frac{\beta(\beta-1)}{2}(\omega -p + p(\omega -p)^2) \sqrt{B_0} r^{\beta/2-2}
\eeqa
Hence we find diverging tides in the asymptotic region for all $\beta>2$.
For $\beta=2$ we find finite tides at infinity; however,
this implies that at infinity the energy density approaches a constant, and
hence the total energy per unit length of this wave diverges.  
Indeed, this might have been expected as these solutions are not
asymptotically flat.
Instead, they represent geometries with gravitational wave 
energy concentrated far away from the black string.

To close this section, we 
will discuss the results for the remaining orthonormal components of the
curvature tensor. It turns out that there are two cases to consider:
$R^{abcd}$ and $R^{a \alpha b \beta}$ where $\alpha, \beta$
take values in $0,4,r$ while the remaining frame indices are $\theta$
or $\phi$. Straightforward evaluation shows 
that the curvature components with an odd
number of angle indices vanish when evaluated on our radial geodesics
which were chosen so that $\prt_\theta F^2_3=0=\prt_\phi F^2_3$.
Further, as the components in eq.~(\ref{curv}), the nonvanishing components
considered here remain finite in the stationary frame (\ref{funf}).
So clearly the boost (\ref{loren}), which is trivial in the angle
directions, cannot introduce any divergences in $\hat{R}^{abcd}$.
Now for the $R^{a \alpha b \beta}$, we have
symbolically $\hat{R}=L^2 R$ upon boosting to the observer's rest frame,
where we have only indicated the non-trivial components of the boost
matrix. Because the $R^{a \alpha b \beta}$ are everywhere finite,
the tidal forces here diverge only as
badly as the leading order divergence in $L^2$. As $r\rightarrow0$, this
is $1/r^{l+2}$ for $\beta=-(l+1)$, and as $r\rightarrow\infty$, this is
$r^l$ for $\beta=l$, just as found above. Note that various cancellations
above reduced the singularities from what one might expect for $L^4$.
Also as above, the original curvature components supply a factor of
$l(l+1)$ multiplying these divergent terms so that the special cases
$\beta=-1,0,1$ survive without any divergent tidal forces.
These qualitative arguments are confirmed with direct evaluation.
Hence one finds that there are no worse divergences
than found by considering only the $0,4,r$ subspace.

Again, our final conclusion is that all of the modes with $\beta=-(l+1)<-1$
in fact produce null singularities at $r=0$, while for $\beta=l>1$, 
singularities appear in the asymptotic region $r\rightarrow\infty$.

\section{Longitudinal and Transverse Oscillations}

We now return to consider 
the case of the longitudinal waves (\ref{sol29}) with $l=0,\beta=-1$
and that of the transverse waves (\ref{sol28}) with $l=1=\beta$.
The preceding analysis revealed no divergent tidal forces in either
of these cases. There is still the possibility that divergent tidals
might occur when the analysis is extended to consider
derivatives of the curvature (much like the case discussed in \cite{W}).
In the case of the transverse waves, the worry would be that problems
arise in the asymptotic region where the Lorentz boost (\ref{loren})
is divergent. However, recall that a coordinate
transformation (\ref{cotrans}) was found for which the metric
became manifestly asymptotically flat.\footnote{Recall that
no divergent tidals were found for $l=0=\beta$ either, but in eq.~(\ref{sol28})
this mode was shown to be purely the result of a coordinate transformation.}
Hence one will never find any divergent tidal forces as $r\rightarrow\infty$.
In fact, for the transversal waves one can argue that the divergence
of the Lorentz transformation (\ref{loren}) is not physical, but comes from
an incorrect choice of gauge in the limit $r\rightarrow \infty$. 
Namely, in this case one could view the asymptotically
flat form of the transverse wave metric (\ref{flat}) with
$\dot A_i = const.$ as the correct example of the leading order behavior 
of slowly oscillating transverse waves. The fact that this solution is
asymptotically flat implies that the Lorentz transformation (\ref{loren}) 
must be finite in the limit $r\rightarrow \infty$.

For the longitudinal waves, the potential problem would be at $r=0$ where
again divergent components appear in the transformation (\ref{loren}).
A definitive demonstration of the regularity of these solutions
would require finding a coordinate transformation for which the
metric (\ref{sol29}) becomes analytic at the null surface.
For a wave profile constant in $u$, we recover the original solution
(\ref{sol11}) with a modified charge in which case
it is straightforward to find the analogue of
Eddington-Finkelstein coordinates, both past and future.
For the nonstationary case, Horowitz and Marolf \cite{HM}
have shown that this solution has a continuous (but not necessarily smooth)
metric at $r=0$. Further,
our theorem shows that all the scalar curvature invariants of this solution
are identical to the original black string, while our preceding analysis
found no evidence of diverging tidal forces for slowly oscillating strings.
In fact, we can make an even stronger statement about the
`invisibility' of the monopole wave: if we compute
the curvature of the oscillating string in an appropriate orthonormal
basis (see below), we find that it does not
depend on the wave at all! 
Given all these results, one is tempted to conjecture that the longitudinal
waves are completely regular, and so we are lead to attempt a
construction of analytic coordinates.
It turns out that we can find such coordinates at
least for a certain class of wave profiles.
A rather surprising result in our analysis
is the quantization of a certain
constant present in the test function, coming from the requirement that all the
derivatives of the metric be continuous
as the null surface is approached. It is not clear to us at present whether
this quantization is an artifact of our ans\"atz or whether it really
represents a physical effect.
We will defer a more detailed investigation of this issue to future work.
At this point, we can only provide some guidelines explaining how we have
arrived at the analytic ans\"atz and the associated quantization condition. 

We begin by considering the longitudinal waves described
by the metric (\ref{sol29}) which may be written:
\beq\label{sol292}
ds^2=\left(1+{Q_1-Q_2+b(u)\over r+Q_2}\right)du^2+{2r\over r+Q_2}du\,dv
+(r+P_1)(r+P_2)\left({dr^2\over r^2}+d\Omega\right)
\eeq
Now the coordinate transformation, $dv=d\hat{v}+{Q_1-Q_2+b(u)\over2Q_2}
du$, simplifies the metric somewhat producing\footnote{If one now constructs
an orthonormal frame analogous to that (\ref{funf}) found for the metric
(\ref{metro2}), one would find that the wave profile $p^2(u)$ completely
disappears from the frame components of the curvature tensor.}
\beq\label{sol293}
ds^2=p^2(u)\,du^2+{2r\over r+Q_2}du\,d\hat{v}
+(r+P_1)(r+P_2)\left({dr^2\over r^2}+d\Omega\right)
\eeq
where 
\beq\label{sol294}
p^2(u)={Q_1+b(u)\over Q_2}\ .
\eeq
Implicitly we assume that $p^2(u)>0$ --- see footnote 6 ---
and as usual $d\Omega$ is the metric on a two-sphere. We  see that
the location of the null surface remains at $r=0$ and that it appears to be
a metric singularity: we know however,
that if $p^2$ were a constant, we could easily show that this singularity is
just a coordinate artifact,
as we have mentioned above. Let us therefore try to follow this argument as
closely as possible. We can
put the ($\hat{v},u$) part of the metric in the form conformal to the
constant $p^2$ case, by defining a new
coordinate $z$: $dz = p^2(u) du$ and $q(z) = 1/p(u)$ in which case the metric
becomes
\beq\label{sol295}
ds^2=q^2(z)\left[dz^2+{2r\over r+Q_2}dz\,d\hat{v}\right]
+(r+P_1)(r+P_2)\left({dr^2\over r^2}+d\Omega\right)\ .
\eeq
Now we can introduce new coordinates
$\tilde v$ and $\tilde z$ in analogy with the tortoise coordinates we
would usually define for a stationary solution. A bit of algebra leads
to the choice
\beqa\label{turtle}
d\tilde{v}&=&d\hat{v}+\frac{P_1+P_2}{2\sqrt{P_1P_2}}
{dr\over r^2}(r+Q_2)(r+\frac{2P_1P_2}{P_1+P_2})
\nonumber\\
d\tilde{z}&=&dz-\sqrt{P_2\over P_1}{dr\over r}(r+P_1)
\eeqa
These coordinates are oriented towards the
future portion of the null surface, and they clearly show the
location of the future horizon as $r\rightarrow 0$ along with
$\hat{v} \rightarrow +\infty$ and $z \rightarrow -\infty$.
Reversing the signs of the shifts to $\hat{v}$ and $z$,
we can go to the past horizon.
These coordinate transformations are designed to `eat up' the
metric singularity manifest in the $r^{-2}$ factor appearing
in $g_{rr}$. If we rewrite the
metric (\ref{sol294}) in terms of the tortoise coordinates
(\ref{turtle}), we find
\beqa\label{quanm2}
ds^2 &=& q^2(z)\left\{ d\tilde{z}^2 + 2 {r\over r+Q_2}d\tilde{z}\,d\tilde{v}
+2\sqrt{P_2\over P_1}{r+P_1\over r+Q_2}dr\,d\tilde{v}
-{P_1-P_2\over\sqrt{P_1P_2}}dr\,d\tilde{z}\right\}
\nonumber\\
&&\qquad
+(r+P_1)(r+P_2)\left[{1-q^2(z)\over r^2}dr^2+d\Omega\right]
\eeqa
Here $z$ is to be understood as an implicit function of $\tilde{z}$ and $r$: 
$z= \tilde{z} + \sqrt{P_2\over P_1}\left[r+P_1\log(r/P_1)\right]$,
and it diverges to
$-\infty$ as $r\rightarrow 0$ as dictated by the logarithm. 
Now if $q^2 = 1$, 
the divergent $g_{rr}$ term in (\ref{quanm2})  would
be absent, and we would obtain a smooth metric at the null surface $r=0$.
Can we now select a wave profile $q^2(z)$ such that a similar cancellation
still occurs in the limit $r\rightarrow 0$?
The answer turns out to be in the affirmative: Consider a
function $q^2(z)$ which
in the limit $z \rightarrow - \infty$ converges to $1 -A^2 
\exp((2+\alpha)z/\sqrt{P_1P_2})$
for some positive $\alpha$.
Substituting the appropriate coordinate transformation $z=z(\tilde{z},r)$
in this expression, we get
that in the limit $r\rightarrow 0$ the wave profile to the leading order is
$q^2 = 1 -A^2 (r/P_1)^{2+\alpha} \e^\fun$, where for convenience
we have defined the linear function $\fun\equiv
(2+\alpha)(\tilde{z}+\sqrt{P_2/P_1}\,r)/\sqrt{P_1P_2}$.
As long as $\alpha\ge 0$, this is
precisely of the form needed to cancel the pole in $g_{rr}$! The metric, in
this limit, becomes
\begin{eqnarray}\label{q3}
ds^2 &=&\left\{1 -A^2 \left({r\over P_1}\right)^{2+\alpha}
\e^\fun\right\}
\nonumber\\
&&\qquad\times\ 
\left\{d\tilde{z}^2 + 2 {r\over r+Q_2}d\tilde{z}\,d\tilde{v}
+2\sqrt{P_2\over P_1}{r+P_1\over r+Q_2}dr\,d\tilde{v}
-{P_1-P_2\over\sqrt{P_1P_2}}dr\,d\tilde{z}\right\}
\nonumber\\
&&\qquad+(r+P_1)(r+P_2)\left[\frac{A^2}{P_1^2} 
\left({r\over P_1}\right)^{\alpha}
\e^\fun dr^2+d\Omega\right]
\end{eqnarray}
which is smooth at $r=0$.

At this point, however, we can push these arguments one
step further. A
glance at the metric (\ref{q3}) immediately shows that all of
its derivatives with respect to the
angles and the coordinates  $\tilde{z}$ and $\tilde t$ 
are well defined as the horizon is approached
--- we ignore subleading terms in $q^2$ for the moment.
Moreover, if $\alpha$ is chosen to be an non-negative integer,
\ie $\alpha=0,1,2,\ldots,$ 
it is easily verified that all the derivatives with respect to $r$
are also well defined in this limit.
Indeed, we see that the only possibly contentious terms in the metric are the
factors $r^{2 + \alpha}$ and $r^{\alpha}$.
If $\alpha$ were not integral with $N > \alpha > N-1$, then
taking $N$ $r$-derivatives of the metric would produce factors which diverge 
as $r\rightarrow0$. At this point, we cannot tell
whether this divergence could cause some covariant derivative of the
curvature tensor to diverge as well (resulting in a null
singularity as in \cite{W}),
or if it could be removed by another even more clever change of coordinates.
On the other hand for integral $\alpha$, we see that after
taking some  number of derivatives of the
metric with respect to $r$, the contentious factors disappear altogether,
leaving an expression which is perfectly well defined 
as $r\rightarrow 0$ and $z\rightarrow-\infty$. 

It is a matter of a simple counting of powers to convince
oneself that all these conclusions
remain unchanged if the function $q^2(z)$ can be written as a uniformly
convergent power series
$q^2(z) = 1 - \sum_{n=2}^{\infty} A^2_n w^n$ for 
$w = \exp(n z/\sqrt{P_1P_2})$ and $z \rightarrow -\infty$.
The subleading contributions would be integer powers of the leading term,
therefore still having well-defined $r$-derivatives. 
The requirement of uniform convergence ensures that
the summations and derivatives commute, 
resulting in the conclusion that the profile $q^2$ is itself analytic
on the horizon.
This allows us to extend our arguments in order to analytically
continue the solution through the past horizon $z\rightarrow \infty$,
as well as the future horizon at $z \rightarrow -\infty$.
For example, consider a wave profile of
the form $q^2(z) = 1 -A^2/\cosh(k z/\sqrt{P_1P_2})$ for some 
positive integer $k\ge2$.\footnote{Note that we would require $A^2<1$
in order to ensure that $p^2=1/q^2>0$, as assumed above.}
Expanding this profile in the limit $z \rightarrow -\infty$
(future horizon), we find  $q^2(z)= 1- A^2 \sum_{m=0}^{\infty}
(-1)^m \exp((2m+1)k z/\sqrt{P_1P_2})$, precisely of the form
guaranteeing the existence of the limit $r \rightarrow 0$. If we approach the
past horizon instead, with
 $z \rightarrow \infty$, we find $q^2(z)= 1- A^2 \sum_{m=0}^{\infty} (-1)^m
\exp(-(2m+1)k z/\sqrt{P_1P_2})$, again
rendering the limit well-defined.
Hence we see that with this example we really have constructed 
a wavy black string with regular future and past horizons.
Of course, this example is easily generalized \eg by taking
linear combinations of inverse $\cosh$'s with different integers $k$,
or by taking a wave profile of the form $A^2/(w^k+w^{-k'})$
with independent $k$ and $k'$.
Hence as we have claimed above, we see that at least for a certain class
of wave profiles, the longitudinal wave solutions
are analytic at $r=0$, which therefore can be identified as a regular
event horizon. Hence we can think of these waves as
time-dependent hair on the black string.
Clearly, our argument is only an existence proof. It would be interesting
to prove that the general
family of longitudinal waves is regular, or to determine the precise
conditions which the wave profile must satisfy in order to ensure regularity.
We note here that for all of our examples above the profiles have essentially
compact support in the $z$ coordinate, and therefore in the horizon limit 
the waves are exponentially damped, the solution approaching the
stationary string.

One could also consider introducing transverse and longitudinal waves 
on the black string at the same time. This would amount to adding to
the line element (\ref{sol292}) a term of the form, \eg
\beq\label{super1}
{r^2 a(u) \cos\theta\over r+Q_2} du^2
\eeq
In this case, we would perform all of 
the same transformations described above with the result that
the final metric (\ref{quanm2}) is modified by the
addition of
\beq\label{zzann}
q^4(z){a(z)\cos\theta\over r+Q_2}\left(r d\tilde{z}+\sqrt{P_2\over P_1}(r+P_1)
dr\right)^2
\eeq
where we have rewritten $a(u)$ as $a(z)$. Now regularity of the horizon
is secured by demanding that $a(z)$ is an
analytic function of $\tilde z$ and $r$. The latter is easily accomplished
with wave forms similar to those discussed for $q(z)^2$ above.

\section{Discussion} 

In this work, we have investigated geometric properties of Garfinkle-Vachaspati
waves, obtained by superposing gravitational oscillations on stationary 
solutions with null hypersurface-orthogonal Killing vectors. We have 
first given a detailed discussion of the adaptation of this solution generating 
technique to a variety of supergravity models, relaxing some of the matter 
sector constraints imposed  previously \cite{GV}. Then we have 
developed a purely geometric theorem, stating that the
GV modes are completely invisible to any scalar invariants
constructed from the metric and matter. 
The proof of our theorem does not rely on the dynamics at all. It
is a consequence of the null symmetry, which must be present in order to 
interpret the solutions as waves, and hence it applies to any metric which
can be represented in the generalized Kerr-Schild form with respect to 
the symmetry. Thus our theorem holds in any theory of gravity and in an 
arbitrary number of dimensions.

As a consequence, it is evident that one must further scrutinize wavy
string geometries using non-invariant probes, such as tidal forces. Recall 
that we have in fact shown that none of the scalar invariants changes 
under the Garfinkle-Vachaspati map. In the group-theoretic language, 
they are invariants of the orbits of the 
solution-generating map. Now, this could suggest that the wavy strings 
might contain curvature singularities identical to those of the 
corresponding stationary solutions belonging to the same orbit. 
Consider, \eg the stationary solutions with all charges equal. Since 
they correspond to the extremal Reissner-Nordstrom black strings,
with traversible horizons and spacelike curvature singularities at the core,
one could be tempted to conjecture that all the GV wavy modes superposed on 
such solutions also have similar curvature singularities - for the curvature 
scalars are exactly the same, and in particular also blow up 
at the `core'. However, in general this is incorrect. The 
argument is incomplete because it does not answer whether such singular 
regions can be causally connected to the physically interesting parts of 
the spacetime, such as the asymptotic infinity. Thus it becomes evident that 
the notion of singularity is a far more complex issue here: in the examples 
of oscillating solutions we have encountered, some of the singularities 
encoded in scalar invariants may not belong to the physical spacetime, because 
no causal geodesics can ever reach it. The simplest example of such behavior 
is given by the solutions discussed by Gibbons, 
Horowitz and Townsend \cite{HGT}. 
These solutions correspond to the choice of the wave degree of freedom $\Psi$ 
such that $f+ b(u)/r=1$ in eq.~(\ref{sol29}). 
Although they belong to the same GV 
orbit as the stationary extremal Reissner-Nordstrom strings, their 
causal structure is dramatically different: in order to get the
analytic extension across the horizon, one has to extricate the region 
normally associated with the `Interior' of the black string
from the physical spacetime, and replace it with a mirror image
of the `Exterior'. The resulting manifold is a completely nonsingular
spacetime with a horizon, with the topology of the Anti-de-Sitter horizon
crossed with a sphere, dividing two asymptotically flat regions. 
Non-curvature singularities, manifest in the presence of geodesic
incompleteness, may reemerge if one attempts to compactify some of the
longitudinal directions in order to descend to lower dimensions.
Nevertheless these singularities are typically milder than the
original ones which got cut out in the process of analytic
extension. Therefore we can see that our theorem states that scalar invariants 
are identical as analytic functions of the coordinates on each GV orbit. The 
observers `surfing' on GV waves of differing geometries however may 
choose to access different regions of this `complex' coordinate plane as 
physical, and hence see different divergences. We believe that this unusual 
behavior is peculiar to wavy strings (or more generally wavy $p$-branes) and 
hence pathological.

With this in hindsight, we have found that most of the undulating
solutions cannot be interpreted as black strings: the excitation of
any localized multipole mode with $l\ge 1$ gives rise to a naked null 
singularity as opposed to an event horizon. Similarly, oscillations of growing 
modes with $l\ge 2$ result in divergences very far away from the string.
To uncover these divergences, we have found a timelike geodesic for each
excited multipole, which extends to infinity and reaches the null surface 
in a finite affine time. We have then propagated an observer along this
geodesic to the vicinity of the region of geometry we wished to explore,
and isolated the leading divergences of the curvature forms
Lorentz-transformed to the rest-frame of the 
infalling observer. These divergences translate into infinite physical
forces detectable by the observer, and hence must be interpreted as naked
singularities. 

One could easily generalize these conclusions to include the waves 
propagating in the remaining `internal' dimensions, which we have ignored 
until now because they were passive spectators for the most part of our 
investigation. In the framework of heterotic string theory this can be done 
as follows: we could take the tensor product of the string frame solution 
(\ref{sol11}) with a four-torus, 
lifting it to nine dimensions. This is guaranteed to be a solution of the 
effective action of heterotic string theory in nine dimensions, with the 
matter fields identical to those of the original five-dimensional solution 
(\ref{sol11}). We could then view the Laplacian 
constraint of (\ref{psiconstr}) as a 
linear combination of the spacetime and internal parts. From
the solution (\ref{sol11}), the Laplacian constraint in the Einstein
frame in nine dimensions can be decomposed as 
$\vec \nabla^2_x \Psi + kl \vec \nabla^2_y \Psi =0$, with 
$k$ and $l$ defined in eq.~(\ref{sol10}). The operators 
$\vec \nabla^2_x$ and $\vec \nabla^2_y$ denote the flat space Laplacians in 
the $3D$ spatial sections and the internal space, respectively. 
Because the internal space is a four-torus, we can expand the wave profile
$\Psi$ in the Fourier series with respect to the internal space basis,
which consists of the exponentials of linear combinations of the internal
coordinates $\exp(\vec c \cdot \vec y)$. The coefficients of the coordinates 
are integer multiples of the inverse radii of the torus, and would give rise to 
the mass of the transversal part of the wave profile $M^2 = \sum^{4}_{i=1} 
c^2_i$. In particular, the zero modes, corresponding to $c_1=\cdots= c_4 =0$
would trivially reduce to the case of 
the massless spacetime waves we have discussed
in the main part of the paper. These are the only scalar harmonics on the 
torus and hence we can ignore them. All the massive modes however turn out to 
be naked singularities. This can be seen as follows: the spacetime part of 
the Laplacian for each massive mode of the wave profile
becomes $\vec \nabla^2_x \Psi_M = M^2 kl \Psi_M$. 
Now using the spherical symmetry of the stationary solution to implement the 
separation of variables $\Psi_M = R_{Ml} Y_{lm}$, we at last obtain the 
equation for the radial degree of freedom: 
\beq \label{radial}
r^2 \ddot R_{Ml} + 2r \dot R_{Ml} - \Bigl( l (l+1) + M^2 P_1P_2 
+ M^2(P_1+P_2) r + M^2 r^2 \Bigr) R_{Ml} = 0
\eeq
From the general theory of differential equations, we know that this 
equation has two classes of solutions - localized (approaching a constant as 
$r\rightarrow \infty$) and growing (diverging in the limit 
$r\rightarrow \infty$). The latter do not represent wavy strings
but rather waves filling up the whole spacetime --- and hence lead to
divergences far away from the string, as in the case of the massless growing
modes discussed in section 4. This can be seen as follows: 
substituting $R_{Ml} = {\cal R}_{Ml}/r$ in (\ref{radial}) and 
keeping only the leading terms in the limit $r\rightarrow \infty$, we find
$\ddot {\cal R}_{Ml} = M^2 {\cal R}_{Ml}$. Hence the growing modes diverge
at infinity as $R_{Ml} \simeq \exp(Mr)/r$ --- \ie faster than any power, and 
so they must produce naked singularities far away from the string for all $l$.
The former case is more interesting, because by the same arguments as above,
far away from the string these solutions damp out as $\exp(-Mr)/r$. 
Thus they might be interpreted as wavy strings --- except that they are 
singular in the limit $r\rightarrow 0$. To show this, we note 
that the point $r=0$ is a regular singular point of the differential equation 
(\ref{radial}). Hence all the solutions admit Laurent series representation, 
with the leading power being $r^{\beta}$. From (\ref{radial}) we find 
$\beta = (-1 \pm \sqrt{1 + 4l(l+1)+4M^2P_1P_2})/2$. The plus 
sign in the definition of $\beta$ leads to finite (in fact, vanishing) limits 
of the radial function as $r\rightarrow 0$ and, as we know from the theory 
of differential equations, it must correspond to the growing solutions which 
are singular at infinity. The minus sign gives solutions which diverge as 
$R_{Ml} \simeq 1/r^{|\beta|}$ on the null surface. In particular, even when
$l=0$, the wave profile diverges with the negative power 
$\beta = -(1+ \sqrt{1+4M^2P_1P_2})/2$. As we have seen in section 4,
such divergences could produce infinite tides --- and since now 
$\beta(\beta+1) = M^2 P_1P_2\ne 0$, the tides we have computed there 
diverge even for the monopole mode! Similarly, we can
see that other localized massive multipole modes also have divergent tides 
on the null surface. Hence we see that all the nontrivial internal modes
lead to naked singularities, which can be detected by tidal forces.

In the case of localized $l=0$ and growing $l=1$ 
modes, we have found a special subclass of wavy strings which 
are analytic on, and everywhere outside of, the null surface, 
and not only continuous - thus extending the earlier results of Horowitz
and Marolf \cite{HM,HM2}. 
A puzzling feature of our special family of analytic black strings 
with non-stationary hair is obviously the quantized nature of the parameter 
$\alpha$. The quantization of $\alpha$ in a technical sense is more
similar to, say, the quantization of the radial modes of an electron in a 
hydrogen atom, where we also require that the wave function of the system is 
analytic --- yet the obvious absence of any quantum-mechanical scales
precludes the possibility that our quantization is of microscopic origin. On
the other hand, it is possible that this is just an artifact of our choice of 
the form of the wave profile yielding analytic metric. In our opinion, this 
ambiguity merits further interest.

Finally, we should remark once more that our analytic examples 
comprise only an existence proof of time-dependent hair. They do not encompass 
the general $l=0$ and $l=1$ modes. Specifically, we do not cover arguably
the most interesting case when the coordinate along the string is
compact, which is a candidate for a hairy black hole in four dimensions. 
Given our demonstration above that the excitation of the passive internal
dimensions leads to naked singularities, we should note here 
that the compact wavy strings may also be singular.
A simple argument supporting this could be based on the fact that 
the compact wave profiles are periodic functions. Hence
the exponential damping of the wave in the horizon limit, manifest in 
eq.~(\ref{q3}) and crucial for demonstrating analyticity of 
the metric in our examples, is absent. This actually turns out to be 
correct --- but to prove that the compact wavy strings are indeed singular
requires a careful examination of the tides beyond the leading order,
and will be given in the forthcoming work of Horowitz and Yang \cite{HY}. 
So, while our examples show that it is possible to have wavy strings 
with regular horizons, and that the waves
can be understood as time-dependent hair on black objects, this identification
must be applied sparingly and discriminatively. In general, the wavy solutions 
are not hairy black strings, or decompactified black holes,
because they contain regions with infinite tides, 
which render the solutions singular. Thus it appears unlikely that the 
classical wave modes alone can give the complete 
accounting of the microscopic black hole degrees of freedom, as proposed
in \cite{LW,Tsey,CT}. The excitation of most of these modes turns the 
original black string into a very bright object. Hence such modes cannot 
be interpreted as black hole hair - in a manner of speaking, they are too 
curly to weave a smooth spacetime. 

\vskip 0.7truecm
\noindent {\bf Acknowledgements}
\vskip 0.7truecm
We would like to thank Gary Horowitz and Don Marolf for helpful conversations.
This work was supported in part by
NSERC of Canada and in part by Fonds FCAR du Quebec.
NK was also supported
in part by an NSERC postdoctoral fellowship.


\begin{thebibliography}{99}

\bibitem{Hair} W. Israel, Phys. Rev. {\bf 164} 1776 (1967);
Comm. Math. Phys. {\bf 8} 245 (1968);
B. Carter, Phys. Rev. Lett. {\bf 26} 331 (1971);
R.M. Wald, Phys. Rev. Lett. {\bf 26} 1653 (1971);
C. Teitelboim, Lett. Nuov. Cimm. {\bf 3} 326 (1972);
Lett. Nuov. Cimm. {\bf 3} 397 (1972);
J. Hartle, Phys. Rev. {\bf D3} 2938 (1971).
J.D. Bekenstein, Phys. Rev. Lett. {\bf 28} 452 (1972);
Phys. Rev. {\bf D5} 1239 (1972);
Phys. Rev. {\bf D5} 2403 (1972).

\bibitem{Entropy} J.D. Bekenstein, Lett. Nuov. Cimm. 
{\bf 4} 737 (1972); Phys. Rev. {\bf D7} 2333 (1973);
Phys. Rev. {\bf D9} 3292 (1974); 
S.W. Hawking, Nature {\bf 248} 30 (1974); 
Comm. Math. Phys. {\bf 43} 199 (1975);
Phys. Rev. {\bf D14} 2460 (1976).

\bibitem{NAb} N. Straumann and Z.-H. Zhou, Phys. Lett. {\bf B243} 33 (1990);
Z.-H. Zhou and N. Straumann, Nucl. Phys. {\bf B360} 180 (1991);
P. Bizon, Phys. Lett. {\bf B259} 53 (1991); 
P. Bizon and R.M. Wald, Phys. Lett. {\bf B267} 173 (1991);
see also T. Torii and K. Maeda,  Phys. Rev {\bf D48} 1643 (1993).

\bibitem{CMP} see for example 
G.W. Gibbons, Nucl. Phys. {\bf B207} 337 ( 1982) 337;
G.W. Gibbons and K Maeda, Nucl. Phys. {\bf B298} 741 (1988);
C.G. Callan, R.C. Myers and M.J. 
Perry, Nucl. Phys. {\bf B311} 673 (1989); 
D. Garfinkle, G.T. Horowitz and A. Strominger,
Phys. Rev. {\bf D43} 3140 (1991);
B.A. Campbell, M.J. Duncan, N. Kaloper  and K.A. Olive,
Phys. Lett. {\bf B251} 34 (1990);
B.A. Campbell, N. Kaloper and K.A. Olive,
Phys. Lett. {\bf B263} 364 (1991);
K. Lee and E.J. Weinberg, Phys. Rev. {\bf D44} 3159 (1991);
A. Shapere, S. Trivedi and F. Wilczek, 
Mod. Phys. Lett. {\bf A6} 2677 (1991);
J.H. Horne and G.T. Horowitz, Phys. Rev. {\bf D46} 1340 (1992);
A. Sen, Phys. Rev. Lett. {\bf 69} 1006 (1992);
S. Coleman, J. Preskill and F. Wilczek,
Phys. Rev. Lett. {\bf 67} 1975 (1991);
Mod. Phys. Lett. {\bf A6} 1631 (1991);
Nucl. Phys. {\bf B378} 175 (1992);
S. Coleman, L.M. Krauss, J. Preskill and F. Wilczek,
Gen. Relativ. Gravit. {\bf 24} 9 (1992).
An elegant account on how these
results fit in with the no-hair theorems
can be found in: J. Bekenstein, e-print gr-qc/9605059.

\bibitem{CMP2} C.G. Callan, J.M. Maldacena and A.W. Peet,
Nucl. Phys. {\bf B475} 645 (1996).

\bibitem{DGHW} A. Dabholkar, J.P. Gauntlett, J.A. Harvey 
and D. Waldram, Nucl. Phys. {\bf B474} 85 (1996)
[hep-th/9511053].

\bibitem{three} W.G. Anderson and N. Kaloper,
Phys. Rev. {\bf D52} 4440 (1995) [hep-th/9503175].

\bibitem{LW} F. Larsen and F. Wilczek, 
Phys. Lett. {\bf B375} 37 (1996) [hep-th/9511064];
Nucl. Phys. {\bf 475} 627 (1996) [hep-th/9604134];
e-print hep-th/9609084.

\bibitem{HM} G.T. Horowitz and D. Marolf, e-prints
hep-th/9605224 and hep-th/9606113.

\bibitem{HM2} G.T. Horowitz and D. Marolf, e-print hep-th/9610171.

\bibitem{Tsey} A.A. Tseytlin, Nucl. Phys. {\bf B477} 431 (1996);
A.A. Tseytlin, Mod.Phys.Lett. {\bf A11} 689 (1996); 
J.G. Russo and A.A. Tseytlin, e-print hep-th/9611047.

\bibitem{Kall} E.A. Bergshoeff, R. Kallosh and T. Ortin,
Phys. Rev. {\bf D47} 5444 (1993); 
Phys. Rev. {\bf D50} 5188 (1994);
E. Bergshoeff, I. Entrop
and R. Kallosh, Phys. Rev. {\bf D49} 6663 (1994);
R. Kallosh, e-print hep-th/9406093.

\bibitem{HT} G.T Horowitz and A.A. Tseytlin,
Phys. Rev. Lett. {\bf 73} 3351 (1994);
Phys. Rev. {\bf D51} 2896 (1995).

\bibitem{CY} M. Cveti\v c and D. Youm,
Phys. Rev. {\bf D53} 584 (1996) [hep-th/9507090].

\bibitem{CT} M. Cveti\v c and A.A. Tseytlin, 
Phys. Lett. {\bf B366} 95 (1996) [hep-th/9510097]; 
Phys. Rev. {\bf D53} 5619 (1996) [hep-th/9512031].

\bibitem{Behr} K. Behrndt, Phys. Lett. {\bf B348} 395 (1995);
Nucl. Phys. {\bf B455} 188 (1995).

\bibitem{GV} D. Garfinkle and T. Vachaspati, 
Phys. Rev. {\bf D42} 1960 (1990); D. Garfinkle,
Phys. Rev. {\bf D41} 1112 (1990).

\bibitem{david} D. Garfinkle, Phys. Rev. {\bf D46} 4286 (1992).

\bibitem{sugra}See for example:\\
A. Salam and E. Sezgin, eds., {\it Supergravity in Diverse Dimensions}
(North-Holland/World Scientific, 1989);\\
M.B. Green, J.H. Schwarz and E. Witten, {\it Superstring Theory},
(Cambridge University Press, 1987).

\bibitem{SH} G.T Horowitz and A.R. Steif, 
Phys. Rev. {\bf D42} 1950 (1990); Phys. Lett. {\bf B258} 91 (1991);
A.R. Steif, Phys. Rev. {\bf D42} 2150 (1990).

\bibitem{S} S.K. Kar, A. Kumar and G. Sengupta.
Phys. Lett. {\bf B375} 121 (1996); A. Kumar and
G. Sengupta, Phys. Rev. {\bf D54} 3976 (1996);
G. Sengupta, e-print hep-th/9609152.

\bibitem{HGT} G.W. Gibbons, G.T. Horowitz and P.K. Townsend,
Class. Quant. Grav. {\bf 12} 297 (1995).

\bibitem{W} D.L. Welch, Phys. Rev. {\bf D52} 985 (1995).

\bibitem{rusty} R.R.~Khuri and R.C.~Myers, Nucl. Phys. {\bf B466} 60 (1996)
[hep-th/9512061].

\bibitem{HY} G.T. Horowitz and H. Yang, in preparation

\end{thebibliography}
\end{document}